%% file: main.tex
\documentclass[]{interact}

\usepackage{epstopdf}% To incorporate .eps illustrations using PDFLaTeX, etc.

\usepackage[numbers,sort&compress]{natbib}% Citation support using natbib.sty
\bibpunct[, ]{[}{]}{,}{n}{,}{,}% Citation support using natbib.sty
\theoremstyle{plain}% Theorem-like structures provided by amsthm.sty

\theoremstyle{definition}

\theoremstyle{remark}

\usepackage{multirow}

\usepackage{siunitx}
\usepackage{caption}
\usepackage{subcaption}
\usepackage{longtable,tabularx}
\usepackage{setspace}
\usepackage{cancel}
\usepackage{float}
\usepackage{multirow}
\usepackage{multicol}
\usepackage{lineno}
\usepackage{tabularx}
\usepackage{array}
\usepackage{pdflscape}
\newcolumntype{Y}{>{\centering\arraybackslash}X}
\begin{document}

% \articletype{ARTICLE TEMPLATE}

\title{Strain-Rate-Consistent $\epsilon$-Based Non-Premixed Flamelet Model}

\author{
\name{Sylvain L. Walsh\textsuperscript{*}\thanks{\textsuperscript{*}Corresponding author. Email: walshsl@uci.edu}, Yalu Zhu, Feng Liu and William A. Sirignano}
\affil{Department of Mechanical and Aerospace Engineering, University of California, Irvine, CA, USA}
}

\maketitle

\begin{abstract}
This numerical study examines a strain-rate inconsistency in the conventional flamelet/progress-variable (FPV) formulation for non-premixed combustion and proposes an alternative coupling based on the turbulence kinetic energy dissipation rate, $\epsilon$. Two-dimensional Reynolds-averaged Navier--Stokes (RANS) simulations of a transonic accelerating reacting mixing layer are performed using one-step kinetics, a conventional FPV model, and the proposed $\epsilon$--$Z$ flamelet model. The analysis focuses on the relation between the RANS-computed mean strain-rate field and the local strain rate imposed on the flamelet through the coupling between the flow computation and the flamelet library. In the FPV formulation, the flamelet state is selected through a transported progress variable, whose evolution is governed by advection, diffusion, and chemical production rather than by the local strain-rate environment. The present results show that this can lead to preferential sampling of near-equilibrium flamelet states in high-strain regions, thereby weakening the intended connection between the computed flow field and the strain-rate-controlled flamelet response. In the $\epsilon$--$Z$ formulation, $\epsilon$ is used to infer the imposed flamelet strain rate, $S^*$, so that the local flamelet state is directly constrained by the modeled turbulence field and the pressure-dependent flammability limit. Selected species are transported explicitly, allowing products to persist through locally quenched regions, while a reactant-availability scaling limits tabulated source terms when the transported composition departs from the flamelet manifold. The results support the use of $\epsilon$ as a physically motivated coupling variable for strain-rate-controlled non-premixed flamelet modelling without introducing an explicit scalar-dissipation-rate closure, while explicit transport of selected species preserves composition transport across locally extinguished regions without relying on a transported progress variable.

\end{abstract}


% \textit{MANUSCRIPT WORD COUNT: 12030 words.}

% \section*{Previous Publication Disclosure}

% This manuscript is a substantially revised and expanded journal article derived from our AIAA SciTech 2026 conference paper \cite{walsh_performance_2026}. The present work is not a direct republication of the conference paper. The model formulation has been physically modified, the manuscript has been completely rewritten, and the analysis has been substantially extended. Most figures have been regenerated. A small number of figures are reproduced or adapted from the conference version. Copyright for the conference paper is held by the authors, as indicated by the copyright statement: “Copyright © 2026 by Sylvain L. Walsh, Yalu Zhu, Feng Liu and William A. Sirignano. Published by the American Institute of Aeronautics and Astronautics, Inc., with permission.”

\input{sections/intro}

\input{sections/numerical_framework}

\input{sections/results}

\section{Conclusions}

% Results obtained with this new flamelet model will be presented in the final version of the manuscript.

Two-dimensional RANS simulations of a transonic, accelerating, non-premixed reacting mixing layer were performed to examine the physical consistency of the conventional FPV model and to assess an alternative $\epsilon$--$Z$ flamelet formulation. The analysis focused on whether the flamelet state selected by the combustion model remains consistent with the local strain-rate environment of the resolved flow.

The FPV results show that the progress-variable formulation can decouple the selected flamelet state from the local strain-rate field. Although the underlying flamelet library is generated from strain-rate-parametrized counterflow solutions, the runtime flamelet state is selected through the transported progress variable. Since this variable evolves through advection, diffusion, and chemical production, rather than through an explicit dependence on local strain rate, the inferred flamelet parameter does not necessarily follow the local turbulent or mean-flow strain-rate field. In the present configuration, this behavior led to preferential sampling of near-equilibrium flamelet states over much of the mixing layer, including regions where the local strain-rate estimate was large. 

Replacing the progress variable with $\epsilon$ as a coupling variable restores a direct connection between the flow-field turbulence state and the flamelet strain-rate coordinate. In the proposed $\epsilon$--$Z$ Flamelet Model, the turbulent kinetic energy dissipation rate is used to infer the imposed flamelet strain rate, $S^*$, which then determines the local flamelet solution. This produces a flamelet response that is more directly tied to the local strain-rate environment and allows strain-induced effects, including flame standoff and local quenching, to appear naturally through the flamelet lookup procedure. The resulting heat-release field is more localized than that obtained with FPV and is concentrated in regions consistent with the pressure-dependent flammability limit of the flamelet library.

The heat-release-rate fields and heat-release-weighted statistics further support this interpretation. Although the FPV and $\epsilon$--$Z$ models use the same underlying flamelet solutions, they sample the library in fundamentally different ways. The FPV model distributes heat release over a broader range of local dissipation states and can assign non-negligible heat release to regions whose inferred strain rate exceeds the local flammability limit. By contrast, the $\epsilon$--$Z$ model restricts chemical activity more directly according to the local value of $S^*$ and $S^*_{\mathrm{fl}}(\bar{p})$, thereby enforcing a closer consistency between the modeled source terms and the strain-rate-controlled extinction behavior of the flamelet library.

The explicit transport of selected species in the $\epsilon$--$Z$ model also addresses a separate limitation of purely table-retrieved strain-rate-based flamelet formulations. When the local strain rate exceeds the flammability limit, the tabulated source terms are set to zero, but products generated upstream remain present through resolved advective and diffusive transport. Thus, the transported composition can remain continuous across locally quenched regions, even though the local chemical source terms are switched off. This permits off-manifold composition states associated with quenching, product transport, and possible downstream re-ignition, without relying on a transported progress variable to impose continuity between reacting and non-reacting flamelet branches. However, because the flamelet library is generated from a single set of scalar boundary conditions, the transported composition need not coincide with the tabulated composition associated with the local values of $Z$, $Z''^2$, $S^*$, and $p$. The proposed reactant-availability scaling provides a pragmatic correction for this mismatch by limiting the tabulated source terms according to the locally transported fuel and oxidizer mass fractions. This allows off-manifold compositions to be treated without introducing additional manifold dimensions for scalar-boundary-condition variations, at the cost of representing these states through a simplified source-term correction rather than a formally expanded flamelet library.

Relative to the conventional FPV model, the proposed formulation increases the number of transported equations because selected species are solved explicitly. However, this additional cost remains lower than that of direct finite-rate chemistry with the full detailed mechanism, and the use of a lumped residual species provides a practical route for reducing the number of transported variables. The results therefore suggest that the added cost may be justified when strain-rate-controlled extinction, localized quenching, and product redistribution through quenched regions are important features of the flow.

Overall, the present results indicate that $\epsilon$ can serve as a physically motivated coupling variable between the resolved turbulent field and subgrid flamelet dynamics in non-premixed flamelet modeling. The proposed $\epsilon$--$Z$ Flamelet Model retains the strain-rate-controlled character of scalar-dissipation-rate-based flamelet approaches while avoiding the need to model a resolved scalar dissipation rate and map it to a stoichiometric flamelet value. At the same time, explicit species transport avoids the abrupt loss of transported products across locally extinguished regions. The present conclusions are limited to a two-dimensional RANS configuration and depend on modeling assumptions in the $\epsilon$--$S^*$ relation, the presumed-PDF treatment, and the selected reduced transported-species set. Future work should therefore examine the sensitivity of the formulation to these closures, extend the model to three-dimensional LES, and assess its predictions against DNS or experimental data.

\section*{Acknowledgement(s)}

 Professor Heinz Pitsch of RWTH Aachen University is acknowledged for providing us access to the FlameMaster code. GPT-5.5 was used solely for language refinement. All data curation, analysis, theoretical modeling, methodology development, computations, interpretation of results, and visualizations are original work conducted by the authors. No AI tool was used to generate scientific content, produce data, perform analysis, develop models, conduct computations, or create figures.

\section*{Disclosure statement}

The authors report there are no competing interests to declare.

\section*{Funding}

The research was supported by the Office of Naval Research through Grant N00014-22-1-2467 with Dr. Steven Martens as program manager.

\bibliographystyle{tfq}
\bibliography{references_zot}

\section*{Notes on contributor(s)}
Sylvain L. Walsh is a Ph.D. student in Mechanical and Aerospace Engineering at the Samueli School of Engineering, University of California, Irvine. His research centers on combustion closure models for RANS and LES simulations. He received his Bachelor’s degree from the Universitat Politècnica de Catalunya in 2021 and his M.S. degree from the University of California, Irvine in 2023.

Yalu Zhu is a postdoctoral researcher in Mechanical and Aerospace Engineering at the Samueli School of Engineering, University of California, Irvine. His research focuses on computational fluid dynamics, turbomachinery, combustion, and the development of numerical methods for reacting-flow simulations. His recent work includes simulations of diffusion flames, reacting turbine stages, turbine-burner configurations, and rocket combustion instability. 

Feng Liu is a Professor of Mechanical and Aerospace Engineering at the Samueli School of Engineering, University of California, Irvine. His research focuses on computational fluid dynamics, turbomachinery, propulsion, aerodynamics, and numerical methods for compressible flows. His work includes multigrid methods for Reynolds-Averaged Navier–Stokes simulations, adaptive grid generation, fluid–structure interaction, unsteady turbomachinery flows, and aerodynamic optimization. He received his B.S. degree from Northwestern Polytechnical University in 1980, his M.S. degree from the Beijing Institute of Aeronautics and Astronautics in 1984, and his Ph.D. degree in Computational Fluid Dynamics from Princeton University in 1991.

William Sirignano is a Distinguished Professor at the Samueli School of Engineering
at the University of California, Irvine. His vast research portfolio over the past 60
years includes work on combustion theory, computational methods, fluid dynamics,
multiphase flows, combustion instability, and propulsion systems. His current research focuses on numerical simulation of practical turbulent reacting flows through Reynolds-Averaged Navier Stokes and Large-Eddy Simulations and high-performance upgrades to gas-turbine engines. He received his M.A. degree in 1962 and Ph.D. degree in 1964, both from Princeton University.

\newpage
\section*{List of Tables}
\begin{enumerate}
    \item Summary of combustion model formulations and associated transport equations.
\end{enumerate}
\section*{List of Figures}
\begin{enumerate}
    \item Flow configuration and computational grid.
    \item Left: Solutions to the flamelet equations presented as S-shaped curves for various background pressures. Right: ${\widetilde{C}_{\mathrm{tab}}({\lambda})=\widetilde{C}({\lambda};\widetilde{Z},\widetilde{Z''^2},\bar{p})}$ relations for different combinations of ${\widetilde{Z}}$ and ${\widetilde{Z''^2}}$ at a pressure of 30 bar.
    \item Mapping of the flamelet maximum temperature and the heat-release rate to the normalized flamelet parameter $\lambda/\lambda_{\mathrm{max}}$ for a background pressure of 30 bar and $\widetilde{Z''^2}\rightarrow0$. 
    \item Strain-rate parametrized flamelet solutions at different background pressures.
    \item Schematic of on- and off- manifold states.
    \item Profiles of temperature and velocity at three different streamwise locations as predicted using the OSK, FPV and $\epsilon$--$Z$ combustion models.
    \item Temperature contours for OSK (top), FPV (center) end $\boldsymbol{\epsilon}$-based (bottom) combustion models.
    \item Left a): contours of the mean strain-rate magnitude, ${S}$. Center b): contours of the flamelet inflow strain rate ${S^*}$. Right c): profiles of ${S}$ and ${S^*}$ at the streamwise location of x = 75 mm (represented by the black vertical lines in a) and b)).
    \item (a) Spatial distribution of the normalized flamelet parameter and (b) Mixture-fraction-integrated progress-variable source term.
    \item Quantities of interest for the FPV results at the streamwise location x = 25 mm.
    \item Heat-release rate, $\widetilde{\dot{{Q}}}$, predicted by the OSK, FPV, and ${\epsilon}$--${Z}$ combustion models.
    \item Heat-release-rate weighted joint statistics of ${\epsilon}$ and $\widetilde{\dot{{Q}}}$ for both flamelet models.
    \item Mixture compositions and ${\widetilde{Y}_{\mathrm{CO}}}$ field.
\end{enumerate}

\end{document}

%% file: sections/intro.tex
\section{Introduction}\label{intro}
In turbulent combusting flows, the most intense chemical activity is often associated with the smallest dynamically relevant turbulent scales. Peak heat release and species production commonly occur at elevated dissipation rates, corresponding to small turbulent length and time scales and, therefore, to large strain rates. Accurately representing this coupling between turbulence and chemistry remains a central challenge in Large Eddy Simulation (LES) and Reynolds-Averaged Navier--Stokes (RANS) computations of reacting flows.

A direct finite-rate chemistry approach, in which chemical source terms are evaluated from Arrhenius-type expressions using the local thermochemical state, is generally too expensive for practical LES and RANS calculations involving detailed reaction mechanisms. The cost arises from the large number of species transport equations and from timestep restrictions imposed by stiff multi-step chemistry. Consequently, reacting-flow simulations often rely on reduced chemistry, such as one-step global mechanisms or small sets of representative reactions. Although computationally efficient, such simplifications may omit important intermediate species and reaction pathways, leading to inaccurate chemical source terms. Moreover, the combustion process is strongly influenced by length and time scales that are not explicitly represented in RANS and may remain subgrid in LES. Therefore, additional modeling is required to represent turbulence--chemistry interaction occurring at the smallest scales.

Flamelet models provide an attractive reduced-order alternative because they treat combustion as a small-scale, locally quasi-laminar and quasi-steady process embedded within the turbulent flow while retaining detailed finite-rate chemistry without transporting all reactive species in the flow-field computation. In the flamelet approach, originally introduced by Spalding~\cite{spalding_mixing_1971} and Bilger~\cite{bilger_structure_1976} and later formalized by Peters~\cite{peters_laminar_1984,peters_turbulent_2000}, turbulent flames are represented as ensembles of locally laminar reaction layers. The flamelet equations are solved independently, often in canonical counterflow configurations, and the resulting thermochemical states are tabulated for use in the flow solver. This permits detailed reaction mechanisms to be incorporated at substantially reduced computational cost~\cite{nguyen_longitudinal_2018}.

In classical mixture-fraction-based flamelet formulations, the flamelet state is typically parameterized by the mixture fraction, $Z$, and by a variable related to the imposed strain rate to the counterflow \cite{peters_laminar_1984,peters_laminar_1988,peters_turbulent_2000}. In the Steady Laminar Flamelet Model (SLFM)~\cite{cook_laminar_1997}, the scalar dissipation rate, $\chi$, at a reference mixture fraction serves as the principal coupling variable between the flow-field computation and the flamelet library. The strain rate (or equivalently, the resulting scalar dissipation rate) governs the flame structure by determining the competition between molecular diffusion and chemical reaction. This formulation preserves a direct connection between the flamelet response and imposed strain rate. However, the SLFM retains only the stable burning branch and the non-reacting branch beyond extinction of the ``s-shaped" curve. As a result, ignition, extinction, and reignition are represented through discontinuous transitions between these branches. In addition, the scalar dissipation rate requires closure both in the turbulent flow field and within the flamelet structure~\cite{peters_laminar_1984,peters_turbulent_2000,claramunt_analysis_2006,poinsot_theoretical_2005}.

The flamelet/progress-variable (FPV) model was introduced for non-premixed combustion to provide a more continuous description of ignition and extinction by incorporating the unstable middle branch of the ``s-shaped" curve, thereby creating a continuous path between reacting and non-reacting flamelet solutions~\cite{pierce_progress-variable_2004}. Initial FPV formulations were derived under the low-Mach-number assumption, which limits their applicability to high-speed compressible flows where thermodynamic nonlinearity, compressibility effects, and viscous dissipation can significantly affect the computed temperature field. To address this limitation, compressible extensions of the FPV framework have been developed. These approaches reconstruct the temperature through analytical expressions obtained from asymptotic expansions of flamelet-scale thermochemical quantities~\cite{pecnik_reynolds-averaged_2012,saghafian_efficient_2015}. An alternative strategy is to retain low-Mach-number flamelet solutions for the mixture composition or chemical production rates, while solving an independent energy equation in the flow field so that the temperature is determined directly by the CFD solution rather than retrieved from the flamelet table \cite{zhan_combustion_2024,walsh_turbulent_2024}. Mixture-fraction-based flamelet formulations, when combined with these compressible FPV methodologies, have become widely used for diffusion-dominated combustion problems. Examples include Refs.~\cite{nguyen_driving_2017,nguyen_impacts_2018,nguyen_longitudinal_2018,nguyen_spontaneous_2019,shadram_neural_2021,shadram_physics-aware_2022,zhan_combustion_2024,walsh_turbulent_2024,pecnik_reynolds-averaged_2012,saghafian_efficient_2015,shan_improved_2021,jiang_species-weighted_2023,coclite_numerical_2015,walsh_turbulent_2025}.

In the FPV framework, the inclusion of the unstable branch is achieved by mapping flamelet solutions originally parameterized by scalar dissipation rate onto a progress variable, $C$. This variable is typically defined as a weighted or unweighted sum of selected product species mass fractions and provides a monotonic coordinate along the ``s-shaped" curve. Coupling to the flow-field computation is then achieved by solving a transport equation for $C$, with a source term obtained from the flamelet table as the sum of species production rates following the definition of $C$. A persistent concern with this formulation is that the transported progress variable evolves through advection, diffusion, and chemical production, without an explicit dependence on the local turbulent strain rate or dissipation rate in the surrounding flow field. Consequently, the computed evolution of $C$ need not reflect the strain-controlled flamelet response that governs extinction, ignition, and local heat release in the canonical flamelet problem.

Recent work on the three-dimensional rotational flamelet model (RFM) has sought to address limitations of standard mixture-fraction-based flamelet approaches by incorporating additional three-dimensional effects, including vortex stretching and centrifugal acceleration~\cite{sirignano_combustion_2021,sirignano_inward_2022,sirignano_stretched_2022,sirignano_three-dimensional_2022,hellwig_three-dimensional_2025,hellwig_vortex_2025,sirignano_flamelet_2026,hellwig_stretched_2026}. Within this framework, Sirignano et al.~\cite{sirignano_flamelet_2026} proposed a flamelet coupling methodology that relates the turbulence kinetic energy dissipation rate, $\epsilon$, obtained from RANS or LES, to the flamelet compressive strain rate, $S^*$. This provides a route for coupling the flamelet state to the local turbulent cascade through a quantity available from the flow-field computation or turbulence model. We do not address vorticity in this present work but note its importance for future study.

The objective of the present study is twofold. First, we examine the behavior of the FPV model with respect to the strain rate experienced by the corresponding flamelet states. In particular, we assess whether the flamelet response implied by the transported progress variable correlates with the local turbulent strain-rate field in the computed flow. Second, we formulate a compressible flamelet model for non-premixed combustion in which the controlling flamelet compressive strain rate is inferred from $\epsilon$. This modified mixture-fraction-based steady flamelet formulation is termed the $\epsilon$--$Z$ flamelet model ($\epsilon$--$Z$ FM). The model restores a strain-rate-controlled flamelet response, as in the SLFM, while avoiding the use of scalar dissipation rate as the primary coupling variable. This work builds on recent preliminary implementations of $\epsilon$-controlled flamelet modeling~\cite{walsh_performance_2026,walsh_flamelet_2026}.

To pursue these objectives, RANS simulations of a a non-premixed flame in a two-dimensional steady transonic accelerating reacting mixing layer are performed using multiple combustion models, including a one-step kinetics (OSK) model, a conventional FPV approach, and the proposed $\epsilon$--$Z$ flamelet model. The simulations are used to compare model behavior, examine the relationship between flamelet state and local turbulent dissipation, and assess the implications of using $\epsilon$ as the controlling parameter for flamelet coupling.

%% file: sections/numerical_framework.tex
\section{Numerical Framework}

\subsection{Flow and Computational Configuration} \label{sec:domain}
To investigate the main objectives outlined in Sec. \ref{intro} while avoiding the complexity of more elaborate geometries and applications, we consider a non-premixed flame in a two-dimensional steady transonic accelerating reacting mixing layer formed by two separate incoming reactant streams with differing compositions, velocities and temperatures. We adopt the configuration originally proposed by Zhu et al. \cite{zhu_numerical_2024}, who studied this flow using a RANS solver with an elliptic formulation, and later used by Walsh et al. \cite{walsh_turbulent_2025}, who employed a RANS solver under the boundary-layer approximation (i.e., a parabolic formulation). In both studies, the mixing layer is subject to a strong favorable pressure gradient of 200 atm/m, imposed to mimic the expanding flow field typical of turbine burner environments. While modeling the turbine burner \cite{sirignano_performance_1999,liu_turbojet_2001} itself is not the aim of the present study, the configuration developed in these works provides a practical setup for our investigation. To produce the desired streamwise pressure gradient, Zhu et al. \cite{zhu_numerical_2024} designed a converging–diverging nozzle geometry.

\begin{figure}{}
\centering
\begin{subfigure}{.55\textwidth}
  \centering
  \includegraphics[width=1.0\textwidth]{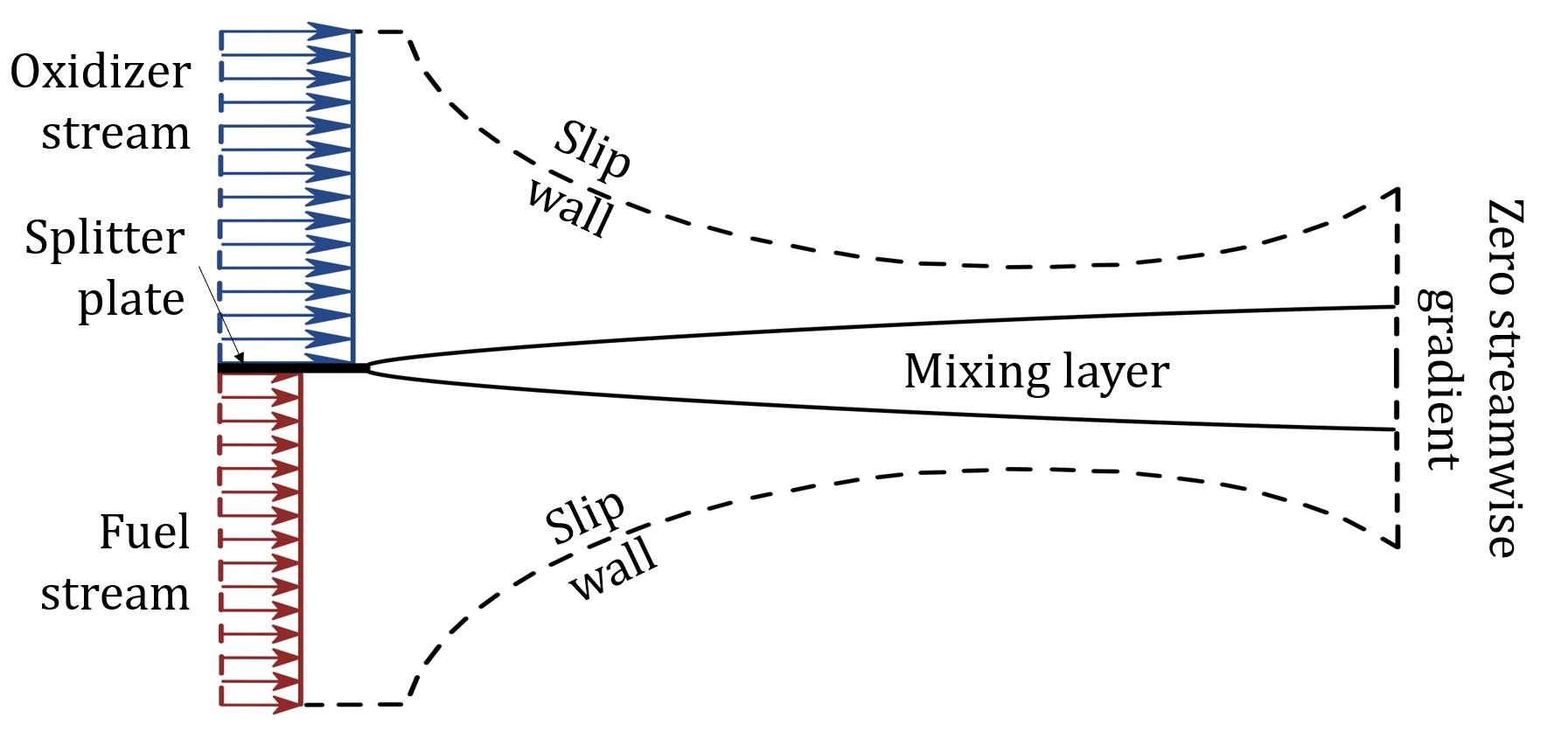}
  \caption{Sketch of the flow configuration.}
  \label{fig:comp_domain}
\end{subfigure}%
\begin{subfigure}{.45\textwidth}
  \centering
  \includegraphics[width=0.97\textwidth]{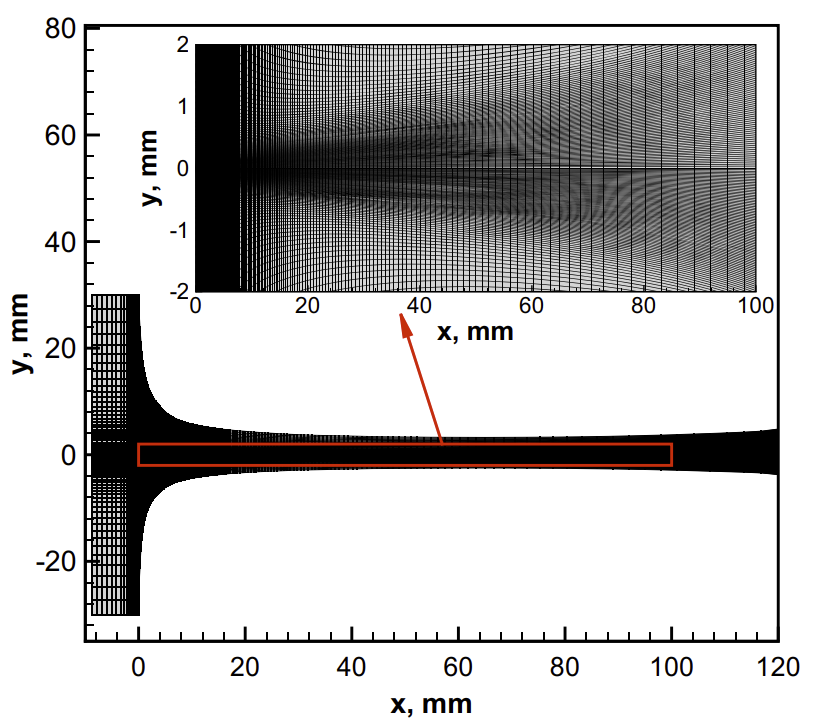}
  \caption{Computational grid.}
  \label{fig:grid}
\end{subfigure}
\caption{Flow configuration and computational grid.}
\label{fig:grid_and_sketch}
\end{figure}

% To produce the desired streamwise pressure gradient, Zhu et al. \cite{zhu_numerical_2024} designed a converging–diverging nozzle geometry. At the inlet, an oxidizer stream enters the upper side of the nozzle and comes into contact with vaporized fuel injected from the lower side. The downstream profiles of the upper and lower walls are determined from the isentropic relations for quasi-one-dimensional flow of air and fuel, based on the inlet flow conditions and initial nozzle height. Since this approach assumes inviscid, non-reacting, and unmixed flow, discrepancies arise between the prescribed pressure distribution and the actual pressure field computed in the reacting mixing layer. These discrepancies are corrected iteratively by reshaping the nozzle contours using isentropic relations to match the -200 atm/m pressure gradient.

Figure \ref{fig:comp_domain} illustrates the converging–diverging nozzle geometry used in this configuration. To minimize the influence of inlet boundary conditions on the developing mixing layer, a uniform inlet section is introduced upstream of the converging zone. The upper and lower nozzle walls are nearly symmetric: both converge rapidly downstream of the inlet, flatten in the middle section, and then gradually diverge past the throat located at $x=70$ mm.

The fuel stream consists of pure gaseous methane (100\% CH$_4$) at a temperature of 400 K, while the oxidizer stream is composed of an equal mass fraction of oxygen and nitrogen (50\% O$_2$, 50\% N$_2)$ at 1650 K. This oxidizer composition differs from that used by Zhu et al.~\cite{zhu_numerical_2024}, who employed air as the oxidizer. The turbulent intensity and the turbulent-to-molecular viscosity ratio at the inlets are set to 5\% and 10, respectively, for the air stream, and to 10\% and 100 for the fuel stream. The static pressure at the inlets is fixed at 30 bar, and the total pressure is specified to yield inlet velocities of 50 m/s for the fuel and 100 m/s for the air.
Inviscid slip with a zero normal pressure gradient and adiabatic wall boundary conditions are specified on the two side surfaces of the nozzle. All flow quantities are extrapolated at the outlet using a zero streamwise gradient, consistent with the supersonic outflow boundary condition, thereby preventing wave reflections into the computational domain. A splitter plate segment is included along the centerline between the inlet and the onset of nozzle convergence to suppress premixing between the fuel and oxidizer streams.

The computational grid used in this study was developed by Zhu et al.~\cite{zhu_numerical_2024} and is shown in Fig.~\ref{fig:grid}, along with a zoomed-in view highlighting local details (top). A curvilinear multiblock structured grid with matched interfaces between neighboring blocks is employed within the nozzle. The vertical grid lines are clustered near the inlet, while the streamwise grid lines are concentrated along the centerline and gradually stretched toward the sidewalls. The ratio of the throat height to the height of the first cell adjacent to the centerline is 440. The grid density was established through a grid-independence study, leading to the selection of a final grid containing 38,272 cells.

\subsection{Governing Equations}
The governing equations required for the compressible, multi-species, reacting RANS computations using all three combustion models are provided hereafter. 
\subsubsection{Reynolds-Averaged Navier-Stokes Equations}
The Reynolds-Averaged Navier-Stokes (RANS) equations for compressible flows with $N$ individual species are expressed by the following transport equations for mass, momentum and energy:
\begin{subequations} \label{eq:navier-stokes}
    \begin{align}
        \frac{\partial\bar{\rho}}{\partial t} + \nabla \cdot (\bar{\rho}\widetilde{\mathbf{V}}) & 
            = 0 \label{eq:continuity} \\
        \frac{\partial(\bar{\rho} \widetilde{\mathbf{V}})}{\partial t} + \nabla \cdot (\bar{\rho}\widetilde{\mathbf{V}} \widetilde{\mathbf{V}}) &
            = -\nabla \bar{p} + \nabla \cdot \boldsymbol{\tau} \label{eq:momentum} \\
        \frac{\partial(\bar{\rho} \widetilde{E})}{\partial t} + \nabla \cdot (\bar{\rho} \widetilde{E}\widetilde{\mathbf{V}}) & = -\nabla \cdot (\bar{p} \widetilde{\mathbf{V}}) + \nabla \cdot (\widetilde{\mathbf{V}} \cdot \boldsymbol{\tau}) - \nabla \cdot \mathbf{q} + \widetilde{\dot Q} \label{eq:energy} 
    \end{align}
\end{subequations}

where $\rho$ is the mixture density, $\mathbf{V}$ is the velocity vector, 
$p$ is the pressure, $\boldsymbol{\tau}$ is the viscous stress tensor, 
$\mathbf{q}$ is the heat-flux vector, and $\dot{Q}$ is the volumetric 
heat-release-rate source term due to combustion. Definitions of 
$\boldsymbol{\tau}$, $\mathbf{q}$, and $\dot{Q}$ are provided below. 
The tilde, $\widetilde{(\cdot)}$, denotes a Favre-averaged quantity, 
whereas the overbar, $\overline{(\cdot)}$, denotes a Reynolds-averaged 
quantity. Here, $\widetilde{E}$ denotes the Favre-averaged total sensible 
energy, given by
\begin{equation}
    \widetilde{E} = \tilde{h} - \frac{\bar{p}}{\bar{\rho}} + \frac{1}{2}\widetilde{\mathbf{V}} \cdot \widetilde{\mathbf{V}}
\end{equation}
with
\begin{equation}
    \tilde{h} = \sum_{n=1}^{N}{\widetilde{Y}_n \tilde{h}_n}\quad\text{and} \quad \tilde{h}_n = \int_{T_{\mathrm{ref}}}^{\widetilde{T}}{C_{p,n}(T)\mathrm{d}T} \:,
\end{equation}
where $C_{p,n}$ is given by NASA polynomials \cite{mcbride_coefficients_1993}, $\widetilde{T}$ is the mean temperature and $\widetilde{Y}_n$ is the mean mass fraction of species $n$. Note that with this definition of enthalpy, the energy transport equation takes a non-conservative form, featuring an explicit mean heat-release rate source term $\widetilde{\dot Q}$, which is provided by the combustion model (as discussed in detail below). 

The perfect gas equation of state is assumed,
\begin{equation}
    \bar{p} = \bar{\rho}R\widetilde{T} \:,
\end{equation}
where $R$ is the gas constant of the mixture, computed by the mass-weighted summation of the gas constant of each species $R_n$ with $R_n = R_0/W_n$, with $R_0$ being the universal gas constant and $W_n$ the molecular weight of species $n$.

\subsubsection{Turbulence Model}
The turbulent viscosity $\mu_T$ is determined by the $k\mbox{-}\omega$ Shear-Stress Transport (SST) model presented by Menter et al. \cite{menter_ten_2003} in 2003:
\begin{subequations} \label{eq:k-omega}
    \begin{align}
        \dfrac{\partial \bar{\rho} k}{\partial t} + \nabla\cdot(\bar{\rho} k \widetilde{\mathbf{V}}) &= P - \beta^*\bar{\rho} k\omega + \nabla \cdot [(\mu+\sigma_k\mu_T) \nabla k] \\
        \dfrac{\partial \bar{\rho} \omega}{\partial t} + \nabla\cdot(\bar{\rho}\omega \widetilde{\mathbf{V}}) &= \dfrac{\gamma\bar{\rho}}{\mu_T}P - \beta\bar{\rho}\omega^2 + \nabla\cdot\left[(\mu+\sigma_\omega\mu_T)\nabla\omega\right] + 2(1-F_1)\dfrac{\bar{\rho}\sigma_{\omega2}}{\omega}\nabla k\cdot \nabla\omega
    \end{align}
\end{subequations}
where $k$ is the turbulent kinetic energy, $\omega$ is the specific turbulent dissipation rate, and $\gamma$, $\beta$ and the $\sigma$ coefficients are model constants. The production term is
\begin{equation} \label{eq:turbulent_Pk}
    P = \mathrm{min}(\mu_TS^2, 10\beta^*\bar{\rho} k\omega) \:,
\end{equation}
where $S = \sqrt{2\mathbf{S}:\mathbf{S}}$ is the magnitude of the strain-rate tensor $\mathbf{S}$, defined below.
The turbulent viscosity is then computed by 
\begin{equation} 
    \mu_T = \dfrac{a_1\bar{\rho} k}{\mathrm{max}\left(a_1 \omega, F_2S\right)} \: .
\end{equation}

The definitions of the blending functions $F_1$ and $F_2$, as well as the model constants, including $a_1$, can be found in Ref. \cite{menter_ten_2003}. $F_1$ approaches to zero away from the wall, and switches to one inside the boundary layer. $F_2$ is unity for boundary-layer flows and zero for free-shear layers. Both of them are artificially set to zero for the turbulent mixing-layer studied here.

\subsubsection{Transport Properties}
The transport terms in the previous equations are given by 

\begin{subequations}
    \begin{align}
        \boldsymbol{\tau} &= 2(\mu + \mu_T) \left[ \mathbf{S} - \frac{1}{3}(\nabla \cdot \widetilde{\mathbf{V}} ) \mathbf{I} \right], 
        \quad 
        \mathbf{S} = \frac{1}{2} \left[\nabla \widetilde{\mathbf{V}} + (\nabla \widetilde{\mathbf{V}})^T \right] \:,\\
        \mathbf{j}_{\phi} &= -\left( \frac{\mu}{\mathrm{Sc}_{\phi}} + \frac{\mu_T}{\mathrm{Sc}_T} \right) \nabla \widetilde{\phi} \:,\\
        \mathbf{q} &= -\left( \frac{\mu}{\mathrm{Pr}} + \frac{\mu_T}{\mathrm{Pr}_T} \right) 
        \left( \nabla \tilde{h} - \sum_{n=1}^{N}{\tilde{h}_n \nabla \widetilde{Y}_n} \right) 
        + \sum_{n=1}^{N}{\tilde{h}_n\mathbf{j}_n} \label{eq:heat_flux} \:.
    \end{align}
\end{subequations}

Here, $\mathbf{j}_{\phi}$ denotes the diffusive flux of the transported scalar $\phi$. For species transport, $\phi=Y_n$, $\mathbf{j}_{\phi}=\mathbf{j}_n$, and $\mathrm{Sc}_{\phi}=\mathrm{Sc}_n$. For the flamelet scalars, $\phi$ represents $\widetilde{Z}$, $\widetilde{Z''^2}$, or $\widetilde{C}$, with corresponding fluxes $\mathbf{j}_z$, $\mathbf{j}_{z''^2}$, and $\mathbf{j}_c$, respectively, and with $\mathrm{Sc}_{\phi}=\mathrm{Sc}$. The molecular viscosity, $\mu$, is computed as the mass-weighted sum of the species molecular viscosities, with each species viscosity evaluated using Sutherland's law~\cite{white_viscous_2005}. In the present work, all species Schmidt numbers are set to unity, $\mathrm{Sc}_n=1$. Consequently, the mixture-averaged Schmidt number is also unity, $\mathrm{Sc}=1$. Under the unity Lewis number assumption, the molecular Prandtl number is therefore $\mathrm{Pr}=1$. The turbulent Schmidt and Prandtl numbers are both taken as 0.7, i.e., $\mathrm{Sc}_T=\mathrm{Pr}_T=0.7$.The last term in Eq.~(\ref{eq:heat_flux}) accounts for energy transport associated with species diffusion, since each diffusing species carries its own specific enthalpy. $\mathbf{I}$ is the Kronecker tensor.

\subsubsection{Species Transport and Flamelet Model Transport Equations}
In this subsection, we summarize the specific transport equations used by each model. 
For the OSK and the $\epsilon$--$Z$ flamelet model, species transport is explicitly resolved through $N$ partial density transport equations:
\begin{equation} \label{eq:species}
    \frac{\partial{\bar{\rho} \widetilde{Y}_n}}{\partial t} + \nabla \cdot (\bar{\rho} \widetilde{Y}_n \widetilde{\mathbf{V}}) = -\nabla \cdot \mathbf{j}_n + \widetilde{\dot{\omega}}_n, \ n = 1, \ 2, \ ..., \ N  \:,
\end{equation}
where $\widetilde{\dot{\omega}}_n$ are the species production rates. 

In contrast, the FPV combustion model eliminates the need for the explicit treatment of the species transport equations. Instead, species mass fractions are obtained directly from a precomputed flamelet library through the transport of the mean mixture fraction $\widetilde{Z}$ and its mean variance $\widetilde{Z^{''2}}$. These quantities are governed by their respective transport equations:

\begin{equation}
\frac{\partial \bar{\rho} \widetilde{Z}}{\partial t} + \nabla\cdot(\bar{\rho} \widetilde{Z}\widetilde{\mathbf{V}}) = -\nabla\cdot\mathbf{j}_z
\label{zequation}
\end{equation}

\begin{equation}
\frac{\partial \bar{\rho} \widetilde{Z^{''2}}}{\partial t} + \nabla\cdot(\bar{\rho} \widetilde{Z^{''2}}\widetilde{\mathbf{V}}) = -\nabla\cdot\mathbf{j}_{z^{''2}}+
2\frac{\mu_{T}}{\mathrm{Sc}_{T}}\nabla\widetilde{Z}\cdot\nabla\widetilde{Z}-\bar{\rho}\widetilde{\chi}
\label{varzequation}
\end{equation}
where the final term in the variance equation represents the mean scalar dissipation rate. This quantity is modeled by relating the integral scalar time-scale and the turbulent flow time-scale, yielding an expression proportional to the turbulent dissipation rate and the mixture-fraction variance:
\begin{equation}\label{eq:mean_chi}
    \widetilde{\chi}=C_{\chi}\frac{\epsilon}{k}\widetilde{Z''^2}
\end{equation}
Here, $\epsilon$ is the turbulence kinetic energy dissipation rate, respectively \cite{peters_turbulent_2000}. $C_{\chi}$ is the proportionality constant relating the turbulent and scalar-mixing timescales. It is commonly assigned the value $C_{\chi}=2.0$ following Janicka and Kollmann~\cite{janicka_two-variables_1979}. However, this value is often adopted as a fixed model constant, with limited case-specific calibration or assessment of its sensitivity.

Additionally, the FPV approach requires an additional non-conserved scalar transport equation to track the value of the mean progress variable $\widetilde{C}$ in the flow field

\begin{equation}
\frac{\partial \bar{\rho} \widetilde{C}}{\partial t} + \nabla\cdot(\bar{\rho} \widetilde{C}\widetilde{\mathbf{V}}) =  -\nabla\cdot\mathbf{j}_c+ \widetilde{\dot{\omega}}_{C} \: ,
\label{cequation}
\end{equation}

\noindent where the source term $\widetilde{\dot{\omega}}_{C}$ is provided by the FPV library. Here, the progress variable represents a composite product-species variable, typically formed as a sum of selected major product mass fractions. Its source term is consequently evaluated as the corresponding sum of the production rates of those species, ensuring consistency between the transported variable and its chemical source term.

\subsection{Numerical Solver}
An in-house three-dimensional code for simulating steady and unsteady transonic flows for single species within turbomachinery blade rows has been developed, validated, and applied by Refs. \cite{zhu_numerical_2017,zhu_flow_2018,zhu_influence_2018,liu_computational_2025}. The code solves the Navier–Stokes equations together with various turbulence models by using the second-order cell-centered finite-volume method based on a multiblock structured grid. The central schemes with artificial viscosity, flux difference splitting schemes, and advection upstream splitting methods with various options to reconstruct the left and right states have been developed and implemented in  the code. 
Recently, it has been extended to include solving transport equations for multiple species with varying specific heat capacities and appropriate chemistry models, and verified and validated by the two-dimensional steady transonic reacting flows in a mixing layer and a turbine cascade \cite{zhu_numerical_2024}, three-dimensional reacting flow in a turbine stage \cite{zhu_large-eddy_2025} and the three-dimensional unsteady reacting flow in a rocket engine \cite{zhu_simulation_2025}. The convective and viscous fluxes are discretized by the Jameson-Schmidt-Turkel scheme \cite{jameson_numerical_1981} and the second-order central scheme, respectively. The local time-stepping method is introduced to accelerate the convergence to a steady state. Thus, the time $t$ in the governing equations is interpreted as a pseudo-time, and a large enough pseudo-time step determined by the local flowfield can be used in each grid cell since time accuracy is not required for steady-state solutions. An operator-splitting scheme is used to treat the stiff chemical source terms (for details see \cite{zhu_numerical_2024}). Parallel techniques based on the message passing interface (MPI) are adopted to further accelerate the computation by distributing grid blocks among CPU processors.

\section{Combustion Models}
In this study, we aim to model methane-air combustion governed by the following global reaction
\begin{equation} \label{eq:methane_reaction}
   \mathrm{CH}_4+2\mathrm{O}_2+2.286\mathrm{N}_2\rightarrow\mathrm{CO}_2+2\mathrm{H}_2\mathrm{O}+2.286\mathrm{N}_2
\end{equation}
Three different combustion models are employed and detailed below. 

\subsection{One-Step Kinetics}
A one-step kinetic combustion model is used as a benchmark case. The chemical reaction rates are determined from mean thermochemical variables using the modified Arrhenius expression:
\begin{equation} \label{eq:Arrhenius_expression}
    {\widetilde{\varepsilon}} = A \widetilde{T}^\beta e^{-E_a/(R_0 \widetilde{T})} \widetilde{[\rm{CH_4}]}^a \widetilde{[\rm{O_2}]}^b
\end{equation}
Here, the concentrations are given by $\widetilde{[X]} = \bar{\rho} \widetilde{Y}_x/W_x$. According to Westbrook and Dryer \cite{westbrook_chemical_1984}, $A = 1.3 \times 10^{9} \, \rm{s^{-1}}$, $\beta = 0$, $E_a = 202.506 \, \rm{kJ/mol}$, $a = -0.3$, and $b = 1.3$ for methane/air combustion. Here, mean species mass fractions for $\mathrm{CH_4}$, $\mathrm{O_2}$, $\mathrm{N_2}$, $\mathrm{CO_2}$ and $\mathrm{H_2O}$ (so $N=5$) are given by the species transport equations (Eq. (\ref{eq:species})). Then, species production rates are determined using stoichiometric coefficients 
\begin{equation}
    \widetilde{\dot\omega}_n = W_n(v_n^{\prime\prime} - v_n^\prime) \widetilde{\varepsilon} \:,
\end{equation}
where $v_n^\prime$ is the stoichiometric coefficient for reactant $n$ in Eq. (\ref{eq:methane_reaction}), and $v_n^{\prime\prime}$ is the stoichiometric coefficient for product $n$.

The mean heat source term (or heat-release rate, HRR) $\widetilde{\dot Q}$ appearing on the right-hand side of the energy equation (Eq. (\ref{eq:energy})) is defined using resolved-scale species production rates,
\begin{equation}
    \widetilde{\dot Q} = -\sum_{n=1}^{N}{\widetilde{\dot\omega}_n h^0_{f,n}} \:.
\end{equation}

\subsection{Mixture-Fraction-Based Flamelet Model} \label{sec:fpv}
 In mixture-fraction-based approaches, such as the SLFM, flamelet libraries are generated for the specific reactant compositions and boundary temperatures described in Sec. \ref{sec:domain} by solving the system of quasi-laminar and quasi-steady flamelet equations in a counterflow configuration~\cite{peters_laminar_1984}, namely,
\begin{equation}\label{eq:steady_flamelet}
-\rho \frac{\chi(Z)}{2}\frac{\partial^2 \psi_j}{\partial Z^2}=\dot{\omega}_j, \hspace{1mm}j=1,2,...,M+1
\end{equation}
Here, $Z$ denotes the mixture fraction and $\psi_j$ represents the reactive scalars, consisting of the mass fractions of the $M$ species considered in the reaction mechanism and temperature. The source terms $\dot{\omega}_j$ correspond to species reaction rates for the mass fraction equations and to the heat-release rate in the energy equation. 

The momentum equations do not appear explicitly in the system above because the scalar dissipation rate, $\chi(Z)$, is prescribed in the following canonical form,
\begin{equation}\label{eq:chi_form}
\chi(Z)=\frac{2S^*}{\pi}\exp{(2\mathrm{erfc}^{-1}(2Z)^2)}\:.
\end{equation}
This formulation is widely adopted in the literature \cite{peters_turbulent_2000,cook_laminar_1997,nguyen_impacts_2018,nguyen_longitudinal_2018,nguyen_spontaneous_2019,shadram_neural_2021,shadram_physics-aware_2022,zhan_combustion_2024,pecnik_reynolds-averaged_2012,saghafian_efficient_2015,shan_improved_2021,jiang_species-weighted_2023,coclite_numerical_2015,walsh_turbulent_2025}, although more comprehensive flamelet formulations that additionally solve the momentum equations momentum equations and properly account for dilatation and vorticity have been proposed \cite{sirignano_three-dimensional_2022,sirignano_inward_2022,hellwig_vortex_2025,hellwig_three-dimensional_2025}. In Eq. (\ref{eq:chi_form}), $S^*$ denotes the imposed normal compressive strain rate at the counterflow inlet. This formulation slightly differs from Peters \cite{peters_turbulent_2000} by a factor of the square root of the ratio of the two densities for the incoming streams, because, we use the strain rate of the incoming fuel stream $S^*$ rather than the oxidizer stream. Notably, $\chi(Z)$ is uniquely parameterized by $S^*$, which in turn is proportional to the stoichiometric scalar dissipation rate $\chi_{st}=\chi(Z_{st})$.

Solving Eq.~\ref{eq:steady_flamelet} for a range of $\chi_{st}$ (or $S^*$) and pressure values, $p$, using a single set of scalar boundary conditions, corresponding to the mixture composition and temperatures of the flow-field simulation reactant streams, yields a steady flamelet library of the form
\begin{equation}\label{eq:psi_chi}
\psi_j=\psi_j(Z,\chi_{st},p)\:,
\end{equation}

\noindent which tabulates the dependence of the reactive scalars on mixture fraction, local strain rate, and background pressure.

These flamelet solutions are computed prior to flow-field simulations using the FlameMaster code \cite{pitsch_flamemaster_2022}. The chemical mechanism employed is a 13-species, 32-reaction skeletal reduction of Version 1.0 of the Foundational Fuel Chemistry Model (FFCM-1) \cite{smith_foundational_2016,tao_critical_2018}, previously used in FPV computations \cite{zhan_combustion_2024,walsh_turbulent_2025}. The species considered in the skeletal reduction are as follows: $\mathrm{H_2}$, $\mathrm{H}$, $\mathrm{O_2}$, $\mathrm{O}$, $\mathrm{OH}$, $\mathrm{HO_2}$, $\mathrm{H_2O}$, $\mathrm{CH_3}$, $\mathrm{CH_4}$, $\mathrm{CO}$, $\mathrm{CO_2}$, $\mathrm{CH_2O}$ and $\mathrm{N_2}$. Nitrogen is treated as inert, meaning it participates in elementary reactions only as a third body. 

Figure~\ref{fig:s-shaped} illustrates the steady flamelet solutions in terms of the ``s-shaped" curve, where the ordinate represents the maximum flame temperature (i.e., the peak value of temperature across mixture fraction $Z$) and the abscissa corresponds to the stoichiometric scalar dissipation rate (which is proportional to the imposed inflow strain rate). Flamelet calculations are performed at background pressures of 5, 10, 15, 20, 25, and 30 bar, representing the range of pressures expected in the flow-field. Each curve exhibits the characteristic bistable structure of non-premixed flames, consisting of an upper stable burning branch and a lower unstable branch, separated by the flammability limit (the point corresponding to the maximum attainable scalar dissipation rate, $\chi_{st,\rm{max}}$, or strain rate, $S^*_{\rm{fl}}$) beyond which the flame quenches. As shown in the figure, the flammability limit increases with increasing pressure due to the higher reaction rates.

\subsection{Flamelet Progress Variable Approach}
At this stage, a progress variable $C$ is introduced to provide a monotonic mapping of the multi-branched flamelet solutions onto a single flamelet parameter, $\lambda$. This parameter is defined as the value of the progress variable at a chosen reference mixture fraction, typically the stoichiometric value, such that $\lambda = C(Z_{st})$. The specific definition of $C$ is somewhat ad hoc and depends on the chemical mechanism employed and boundary thermochemical conditions. However, the formulation requires that the relation between $C$ and $\lambda$ be bijective for all combinations of $Z$ and $p$. For fixed mixture fraction and pressure, the progress variable may be written as $C(\lambda;Z,p)$, defining a one-dimensional dependence on the flamelet parameter. This relation must be invertible so that a prescribed value of $C$ uniquely determines $\lambda$ and, consequently, the corresponding value of $\chi_{st}$. This condition guarantees that every combination of $C$, $Z$, and $p$ maps to a unique flamelet solution. Accordingly, Eq. (\ref{eq:psi_chi}) may be re-described as
\begin{equation}\label{eq:psi_lam}
\psi_j = \psi_j(Z,\lambda,p)
\end{equation}

\noindent where one of the $\psi_j$ is the progress variable $C=C(Z,\lambda,p)$. In the present study, the progress variable is defined as
\begin{equation}\label{eq:c_definition}
C = Y_{\mathrm{H_2O}} + Y_{\mathrm{CO_2}} + Y_{\mathrm{CO}}.
\end{equation}

\begin{figure}[]
\centering
\begin{subfigure}{.45\textwidth}
  \centering
  \includegraphics[width=1.0\textwidth]{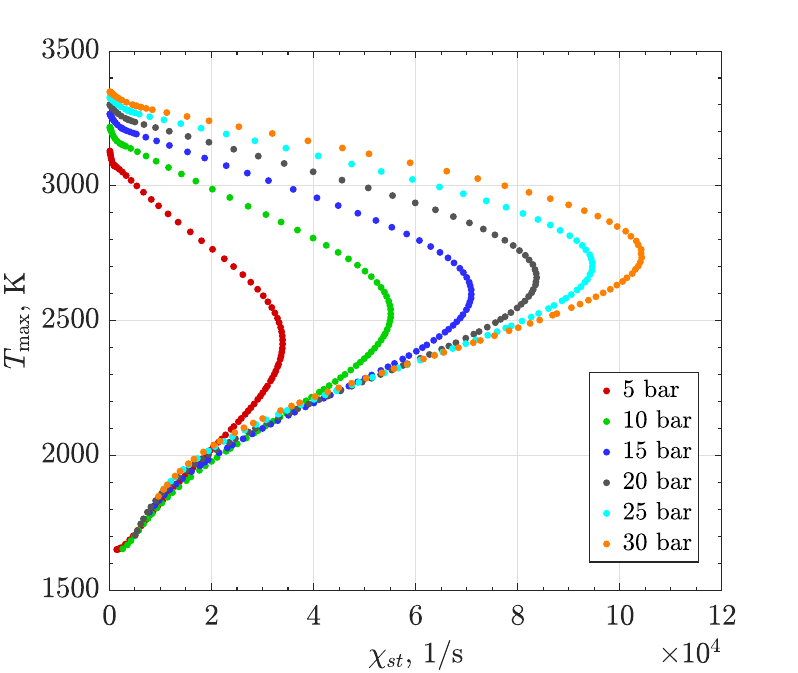}
  \caption{S-shaped curves.}
  \label{fig:s-shaped}
\end{subfigure}%
\begin{subfigure}{.45\textwidth}
  \centering
  \includegraphics[width=0.97\textwidth]{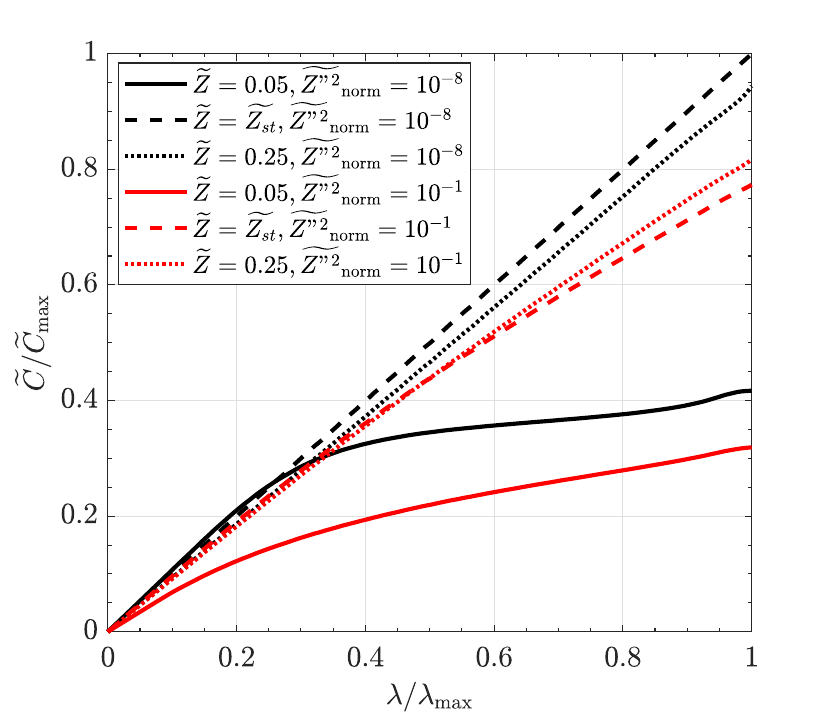}
  \caption{${\widetilde{C}_{\mathrm{tab}}({\lambda})=\widetilde{C}({\lambda};\widetilde{Z},\widetilde{Z''^2},\bar{p})}$ relations.}
  \label{fig:c_lambda}
\end{subfigure}
\caption{Left: Solutions to the flamelet equations presented as S-shaped curves for various background pressures. Right: ${\widetilde{C}_{\mathrm{tab}}({\lambda})=\widetilde{C}({\lambda};\widetilde{Z},\widetilde{Z''^2},\bar{p})}$ relations for different combinations of ${\widetilde{Z}}$ and ${\widetilde{Z''^2}}$ at a pressure of 30 bar.}
\label{fig:s-and-pv}
\end{figure}

The choice of major products in defining C guarantees the monotonic behavior of $\lambda$ along the ``s-shaped" curves of Fig. \ref{fig:s-shaped}. $\lambda_{\max}$ thereby occurs at the largest value of $T_{\max}$. Although this definition may be considered conventional, more systematic strategies for selecting the progress variable and its weights have been proposed in the literature~\cite{ihme_regularization_2012,najafi-yazdi_systematic_2012,bojko_formulation_2016}. Nonetheless, the findings of the present study are believed to be insensitive to the specific choice of $C$. 

Once the laminar flamelet solutions are mapped to the progress variable, they are convoluted to obtain Favre-averaged first moments of the reactive scalars using the presumed probability density function (PDF) approach,

\begin{equation}\label{eq:lambda_mean}
\widetilde{\psi}_j(\widetilde{Z},\widetilde{Z''^2},\lambda,\bar{p})=\int_0^1 \psi_j(Z,\lambda,\bar{p})\widetilde{P}(Z,\widetilde{Z},\widetilde{Z''^2})dZ
\end{equation}

In this work, a standard $\beta$-PDF is used for the mixture fraction $Z$, and a Dirac $\delta$-PDF is assumed for the flamelet parameter $\lambda$. For further implementation details, the reader is referred to \cite{walsh_turbulent_2025}. Note that the reactive scalars still remain functions of $\lambda$. The final step in the FPV formulation consists of inverting the bijective relation $\widetilde{C}_{\mathrm{tab}}(\lambda)=\widetilde{C}(\lambda;\widetilde{Z},\widetilde{Z''^2},\bar{p})$. These relations are computed by Eq. (\ref{eq:lambda_mean}), and the subscript "tab" is used to distinguish the precomputed tabulated value of the mean progress variable from the value obtained during the flow-field computation via its transport equation. This inversion defines a one-to-one mapping between the quantities $\{\widetilde{Z}, \widetilde{Z''^2}, \widetilde{C}_{\mathrm{tab}}, \bar{p}\}$ and a value $\lambda$ and its associated $\chi_{st}$ value. In other words, each combination of $\widetilde{Z}$, $\widetilde{Z''^2}$, $\widetilde{C}_{\mathrm{tab}}$ and $\bar{p}$ uniquely identifies a single steady flamelet solution of Eq. (\ref{eq:steady_flamelet}), i.e., $\widetilde{\psi}_j = \widetilde{\psi}_j(\widetilde{Z},\widetilde{Z''^2},\widetilde{C}_{\mathrm{tab}},\bar{p})$, corresponding to a certain value of $\chi_{st}$ or $S^*$.
Figure \ref{fig:c_lambda} shows the bijective relation $\widetilde{C}_{\mathrm{tab}}(\lambda)=\widetilde{C} (\lambda;\widetilde{Z},\widetilde{Z''^2},\bar{p})$ for various combinations of $\widetilde{Z}$ and $\widetilde{Z''^2}$ at a pressure of 30 bar. The bijectivity of this mapping has ben ensured by the present choice of the progress-variable definition (Eq. (\ref{eq:c_definition})).

Typically, this inversion is performed offline prior to the CFD simulation to reduce computational cost and the flamelet tables are stored as  $\widetilde{\psi}_j = \widetilde{\psi}_j(\widetilde{Z},\widetilde{Z''^2},\widetilde{C}_{\mathrm{tab}},\bar{p})$. However, in the present work, the inversion is evaluated online during the flow-field computations, allowing for the local $\lambda$ values to be stored for analysis. This requires the tabulation of $\widetilde{C}_{\mathrm{tab}}(\lambda)=\widetilde{C} (\widetilde{Z},\widetilde{Z''^2},\lambda,\bar{p})$, which can be stored as an additional reactive scalar within the flamelet libraries, defined as
\begin{equation}\label{eq:mean_tables}
\widetilde{\psi}_j = \widetilde{\psi}_j(\widetilde{Z},\widetilde{Z''^2},\lambda,\bar{p}).
\end{equation}

At runtime, $\widetilde{Z}$, $\widetilde{Z''^2}$, $\widetilde{C}$ and $\bar{p}$ are obtained by solving their respective flow-field transport equations. The Favre-averaged reactive scalars are then retrieved by first determining the flamelet parameter $\lambda$, using $\widetilde{C}$ as the target in the inversion of $\widetilde{C}_{\mathrm{tab}}(\lambda)$. Once $\lambda(\widetilde{Z},\widetilde{Z''^2},\widetilde{C},\bar{p})$ is obtained, the corresponding reactive scalars are evaluated through quadrilinear interpolation of the stored flamelet tables defined by Eq. (\ref{eq:mean_tables}). Consequently, the retrieved $\widetilde{Y}_n$ ($n=1,\ldots,N$) are used to evaluate thermodynamic and transport properties, while the retrieved source terms $\widetilde{\dot{Q}}$ and $\widetilde{\dot{\omega}}_C$ are provided to the energy equation (\ref{eq:energy}) and progress-variable equation (\ref{cequation}), respectively. Note that in the FPV approach, $N=M$. 

Now, we compare compare Eq. (\ref{eq:mean_tables}), representing the flamelet solutions parameterized by the flamelet parameter $\lambda$, with Eq. (\ref{eq:psi_chi}) where the flamelet solutions are parameterized by the stoichiometric scalar dissipation rate $\chi_{st}$ (i.e., the strain rate). In the latter formulation, the strain rate naturally arises from the system of flamelet equations and serves as the primary mechanism governing the balance between diffusive transport and chemical reaction rates, thereby determining the flame structure. In contrast, the progress-variable formulation replaces this physically grounded parameterization with one based on $\widetilde{C}$ (or more precisely, $\widetilde{C} \rightarrow \lambda$), which is chemically driven through the source term $\widetilde{\dot{\omega}}_C$ and lacks any explicit dependence on the strain rate. 

\begin{figure}
    \centering
    \includegraphics[width=0.7\linewidth]{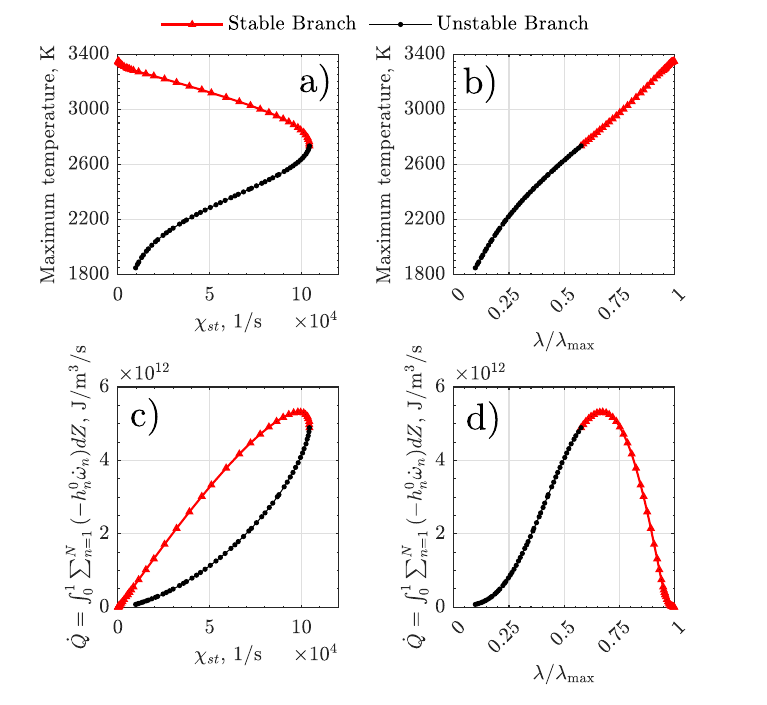}
    \caption{Mapping of the flamelet maximum temperature and the heat-release rate to the normalized flamelet parameter $\lambda/\lambda_{\mathrm{max}}$ for a background pressure of 30 bar and $\widetilde{Z''^2}\rightarrow0$. }
    \label{fig:qmapping}
\end{figure}

Figure ~\ref{fig:qmapping} shows the mapping of the maximum temperature and the heat-release rate to the normalized flamelet parameter in the limit of $\widetilde{Z''^2}\rightarrow0$ and at a background pressure of 30 bar; however, the analysis is unchanged for other combinations of pressure and mixture-fraction variance. The top panels of Fig.~\ref{fig:qmapping} illustrate this transformation in terms of maximum temperature, showing the flamelet solutions parameterized by $\chi_{st}$ as an ``s-shaped" curve in (a), and their corresponding representations parameterized by the normalized flamelet parameter $\lambda/\lambda_{\mathrm{max}}$ in (b). The bottom panels show the corresponding mapping of the integrated burning rate, obtained by integrating the heat-release rate $\dot{Q}$ over the mixture fraction coordinate $Z$: the distribution in terms of $\chi_{st}$ is shown in (c), and the same quantity expressed in terms of $\lambda/\lambda_{\mathrm{max}}$ is shown in (d).

Although the local heat-release rate communicated to the resolved scale originates from a single value of $\widetilde{Z}$ at each grid point, the integrated heat-release rate provides the relative burning strength of the flamelet across the strain-rate spectrum. As shown, the integrated burning rate increases approximately linearly with strain rate, reaching its maximum just before the extinction limit. For each strain rate below extinction, two distinct flamelet solutions exist, corresponding to the upper (stable) and lower (unstable) branches of the ``s-shaped" curve. However, this multiplicity is lost when the solutions are mapped to the progress-variable space via $\lambda$. The resulting profile assumes a Gaussian-like shape, peaking at the value of $\lambda$ corresponding to the flammability-limit strain rate and vanishing at both the lower and upper bounds of the progress variable. 
For the FPV model to remain physically consistent in terms of strain rate, the flamelet parameter $\lambda$ would need to exhibit behavior that correlates with the local strain rate. However, as will be demonstrated in Sec. \ref{sec:results}, this correspondence does not hold in practice.

\subsection{$\boldsymbol{\epsilon}$--$\mathbf{Z}$ Flamelet Model}
In this section, the formulation of the proposed $\epsilon$--$Z$ flamelet model is presented. The objective is to relate the flamelet strain-rate environment to the turbulence cascade process, which determines the smallest dynamically relevant length and time scales at which the flamelet exists. This is accomplished by using the turbulence kinetic energy dissipation rate, $\epsilon$, obtained from the flow-field transport equations, to infer the local strain rate imposed on the flamelet. The formulation follows the theoretical framework proposed by Sirignano et al.~\cite{sirignano_flamelet_2026}. In that framework, the viscous dissipation rate, $\Phi$, for a three-dimensional rotating counterflow is related to the imposed normal compressive strain rate at the inflow, $S^*$, and by the normalized major and intermediate tensile strain rates, $S_1$ and $S_2$, respectively:

\begin{equation}\label{eq:viscous_dissipation}
\epsilon = \frac{\Phi}{\rho} = 2 \nu[(S^*S_1)^2 + S^{*2} + (S^*S_2)^2] 
= 4 \nu S^{*2} [ S_1^2 + 1 - S_1]   
\end{equation}

\noindent with $S_1+S_2=1$. This results in the following relationship between $S^*$ and $\epsilon$: 

\begin{eqnarray}\label{eq:sstar-eps}
S^* = \frac{1}{2}\sqrt{ \frac{C_{vd}\ \epsilon}{ \nu[S_1^2 + 1 - S_1]} }
\end{eqnarray}

\noindent where $C_{vd}$ is a dimensionless coefficient accounting for the distribution of viscous dissipation across a range of the smallest turbulent scales. For the counterflow solution to be physically valid, the constraint $C_{vd} < 1$ must be satisfied (see \cite{sirignano_flamelet_2026}). While this coefficient remains to be definitively determined from direct numerical simulation (DNS) data, a value of $C_{vd} = 1$ is adopted in the present study, consistent with the assumptions outlined in \cite{sirignano_flamelet_2026}. Furthermore, $S_1$ is set to $1/2$, corresponding to an axisymmetric counterflow configuration, commonly used in the SLFM and FPV approaches. The molecular kinematic viscosity, $\nu$, is obtained from the CFD flow-field transport properties. The turbulent kinetic energy dissipation rate, $\epsilon$, is evaluated from the turbulence quantities $k$ and $\omega$ using the standard relation
\begin{equation}
    \epsilon = C_{\mu} k\omega \:,
\end{equation}

\noindent where $C_{\mu} = 0.09$ is a model constant specified by the $k-\epsilon$ turbulence model \cite{chien_predictions_1982}. With this formulation, for any given spatial and temporal location in the CFD domain, $\epsilon$ is used to infer the corresponding flamelet-scale strain rate in accordance with turbulence cascade scaling. This approach reflects the well-established physical principle that scalar gradients intensify as the turbulent length scale decreases, enabling a consistent coupling between turbulence structure and flamelet dynamics. Given that $S^*$ is now known, the solutions of the flamelet system, Eq. (\ref{eq:steady_flamelet}) presented in Sec. \ref{sec:fpv} may be parametrized in terms of $S^*$ as
\begin{eqnarray}
\psi_j = \psi_j(Z,S^*(\epsilon/\nu),p)\:.
\end{eqnarray}

\begin{figure}[]
\centering
\begin{subfigure}{.45\textwidth}
  \centering
  \includegraphics[width=1.0\textwidth]{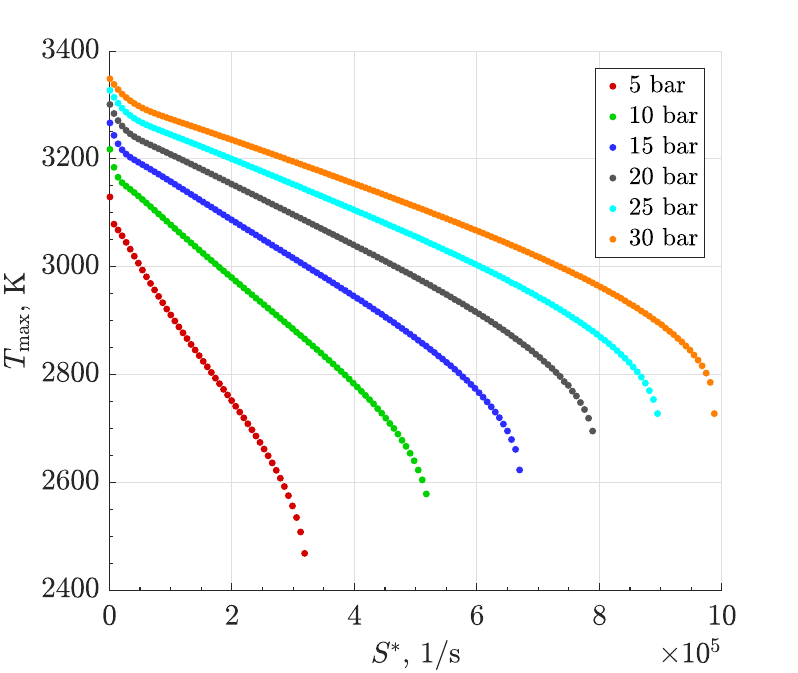}
  \caption{S-shaped curves parametrized by $S^*$.}
  \label{fig:tmax_sstar}
\end{subfigure}%
\begin{subfigure}{.45\textwidth}
  \centering
  \includegraphics[width=1.0\textwidth]{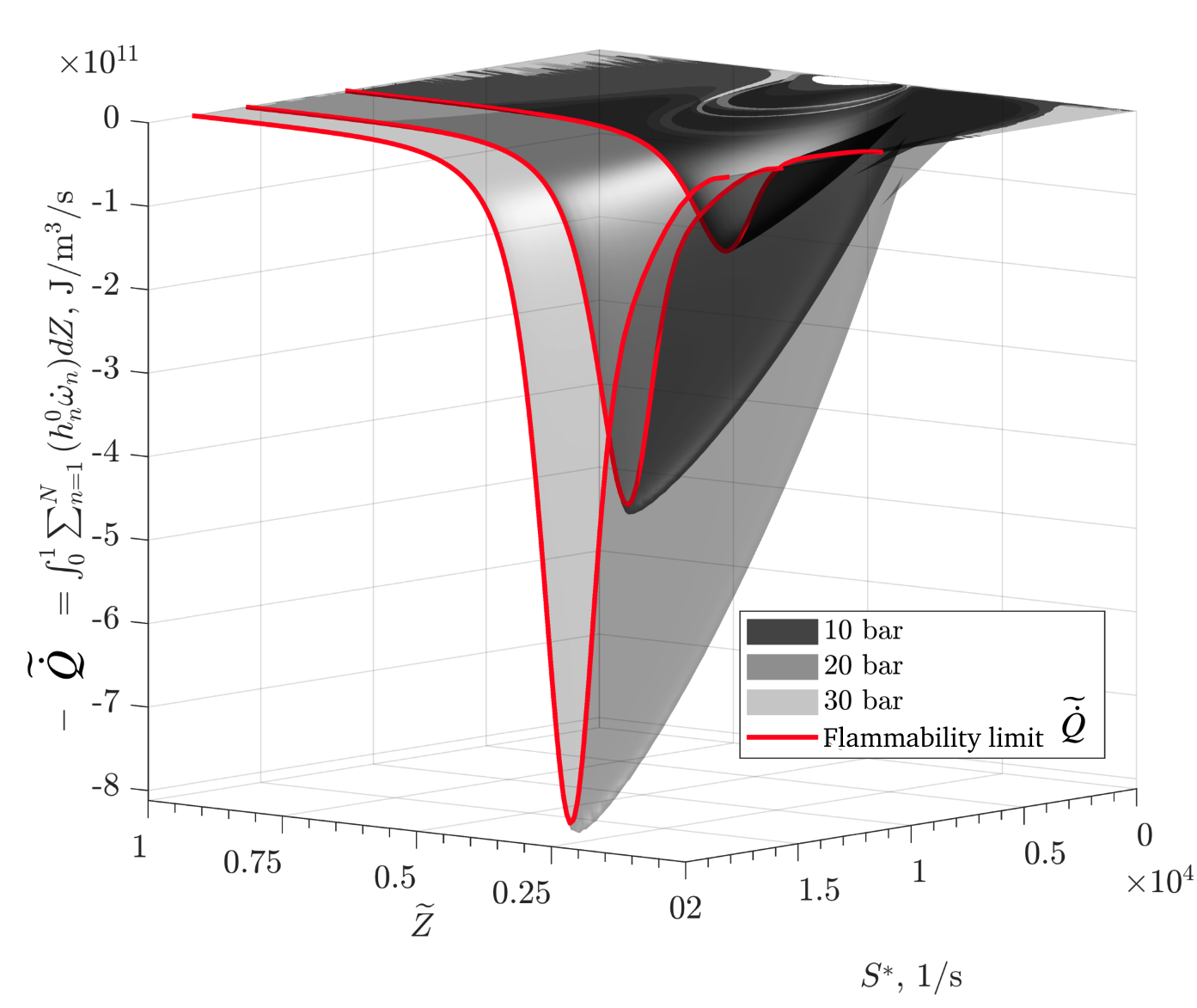}
  \caption{Heat-release rate surfaces as functions of $S^*$ and $\widetilde{Z}$.}
  \label{fig:q_sstar}
\end{subfigure}
\caption{Strain-rate parametrized flamelet solutions at different background pressures.}
\label{fig:sstar_and_q}
\end{figure}

Figure~\ref{fig:tmax_sstar} presents the same set of flamelet solutions previously shown in Fig.~\ref{fig:s-shaped}, but now parameterized by $S^*$
instead of $\chi_{st}$. Notably, as in the SLFM, the unstable branch of the classical S-shaped curve is absent; only the stable branch extending from $S^*\rightarrow0$ to the flammability limit $S^*_{\rm{fl}}$ remains. This occurs because $S^*$ alone cannot provide a bijective mapping of both the stable and unstable solutions. As observed previously, the flammability limit increases with pressure due to the higher chemical rates. Figure~\ref{fig:q_sstar} shoes surfaces of $-\widetilde{\dot{Q}}$, in units of energy per unit volume per unit time, as functions of $S^*$ and $Z$ for different background pressures. The local volumetric heat-release rate increases with increasing strain rate, even though the corresponding residence time decreases. This apparent discrepancy is a consequence of the spatial compression of the flamelet structure: larger strain rates reduce the physical thickness of the reaction zone, thereby concentrating the heat release within a smaller volume. Therefore, the volumetric heat-release rate increases with $S^*$. At sufficiently large $S^*$, however, the reduced residence time limits chemical conversion, leading to extinction. Beyond the flammability limit strain rate $S^*_{\mathrm{fl}}$, indicated by the red lines, the volumetric heat-release (and species production) rates go to zero.

To obtain the mean reactive scalars for use in the flow-field computations, the PDF approach used in the FPV framework is applied here as well. The Favre-averaged reactive scalars are obtained through the convolution given in Eq. (\ref{eq:lambda_mean}). Here, we employ the standard $\beta$-PDF for the mixture fraction $Z$, while for $\epsilon$ a Dirac $\delta$-PDF is used, reflecting the current absence of a more appropriate statistical description. This assumption may warrant further investigation.

\begin{equation}\label{eq:mean_tables_eps}
\widetilde{\psi}_j = \widetilde{\psi}_j(\widetilde{Z},\widetilde{Z''^2},S^*,\bar{p}).
\end{equation}

During runtime, combinations of the resolved-scale quantities $\widetilde{Z}$, $\widetilde{Z''^2}$, $S^*$ and $\bar{p}$ (where $\widetilde{Z}$ and $\widetilde{Z''^2}$ are obtained by the same transport equations utilized in the FPV approach, i.e., Eqs. (\ref{zequation}) and (\ref{varzequation})), determine the local flamelet state. The corresponding reactive scalars, $\psi_j$, are then retrieved through quadrilinear interpolation of the precomputed tables defined by Eq. (\ref{eq:mean_tables_eps}). If the local strain rate exceeds the flammability limit, the table returns a non-reacting (extinguished) solution.

Unlike the FPV framework, where the reaction progress evolves according to the transport equation for $C$, the present $\epsilon$--$Z$
formulation inherently allows for quenching and re-ignition through its dependence on the dynamically varying local strain rate. Accordingly,
$\epsilon$ should not be interpreted as a progress variable, since its magnitude may increase or decrease in both space and time. Rather, it serves as a coupling variable, linking the resolved-scale turbulence characteristics to the subgrid flamelet dynamics.

\subsubsection{Comparison to $\chi$-Based Models}

At first sight, the present formulation resembles a scalar-dissipation-rate-parameterized flamelet model, such as the SLFM, because both approaches ultimately describe flamelet response in terms of the strain imposed on the flamelet. However, the coupling strategy is different. In conventional $\chi$-based models, the flamelet library is parameterized by the stoichiometric scalar dissipation rate, $\chi_{st}$, whereas the flow solver provides, or models, a mean scalar dissipation rate, $\widetilde{\chi}$. These two quantities are not identical and must be related through an assumed scalar-dissipation-rate profile across mixture-fraction space.

In RANS, the mean scalar dissipation rate is commonly modeled following Eq.~(\ref{eq:mean_chi})~\cite{peters_turbulent_2000}. This closure requires specification of the model constant $C_{\chi}$, and the sensitivity of flamelet predictions to this constant has been noted previously~\cite{cao_effect_2007}. In LES, the mean or filtered scalar dissipation rate is often modeled using the first-order gradient expression of Girimaji and Zhou~\cite{girimaji_analysis_1996},
\begin{equation}
\widetilde{\chi}
=
\left(D_Z+D_T\right)
\left(\nabla \widetilde{Z}\right)^2 ,
\label{eq:mean_chi_les_text}
\end{equation}
where $D_Z$ is the molecular diffusivity of the mixture fraction and $D_T$ is the turbulent diffusivity. Examples of this LES treatment include Refs.~\cite{pitsch_scalar_2000,kemenov_modelling_2012}.

The modeled flow-field scalar dissipation rate must then be mapped to the stoichiometric scalar dissipation rate required by the flamelet library. This mapping is obtained by invoking the assumed flamelet scalar-dissipation-rate profile, Eq.~(\ref{eq:chi_form}), from which the corresponding stoichiometric value inferred from the modeled flow-field quantity is
\begin{equation}
\widetilde{\chi}_{st}
=
\widetilde{\chi}(\widetilde{Z})
\frac{f(Z_{st})}{f(\widetilde{Z})}.
\end{equation}
Thus, in conventional $\chi$-based coupling, a scalar dissipation rate is first modeled from mean or filtered flow-field quantities and is then projected onto the assumed flamelet profile to recover the library coordinate $\widetilde{\chi}_{st}$.

In the present $\epsilon$--$Z$ model, the assumed functional form for $\chi(Z)$ in Eq.~(\ref{eq:chi_form}) is still used during generation of the flamelet solutions, as in the SLFM, because the flamelet equations are solved without the momentum equations. However, this internal flamelet profile is not used to couple a modeled flow-field value of $\widetilde{\chi}$ to the library. Instead, the imposed flamelet strain-rate parameter, $S^*$, is obtained directly from the turbulent kinetic energy dissipation rate, $\epsilon$, which is already available from the turbulence closure in RANS or LES.

The distinction is therefore one of coupling. In a conventional $\chi$-based approach, the strain-rate information reaches the flamelet indirectly through a modeled $\widetilde{\chi}$ field and an assumed mapping from $\widetilde{\chi}$ to $\widetilde{\chi}_{st}$. In the present formulation, the flamelet is coupled directly through the imposed strain-rate parameter $S^*$. The scalar-dissipation-rate profile within the flamelet then follows from the flamelet solution associated with that imposed strain rate, rather than from an additional model for the flow-field scalar dissipation rate. In this sense, the present approach avoids the intermediate closure for $\widetilde{\chi}$ and anchors the flamelet state directly to the turbulence cascade through $\epsilon$.

\subsubsection{Treatment of the Flammability Limit Discontinuity}
However, reverting to a strain-rate-controlled flamelet formulation reintroduces an inconsistency that FPV models were designed to avoid. When the local strain rate exceeds the flammability limit, the flamelet tables return zero product mass fractions, corresponding to complete local extinction. In formulation where the mixture composition is retrieved from the tables, products generated upstream would therefore not persist through locally quenched regions unless an additional transport mechanism is provided. In the FPV models, this artifact is mitigated because the resolved progress-variable equation provides a transported memory of upstream product formation.

To retain this transport-history effect in the proposed $\epsilon$--$Z$ formulation, partial density transport equations, Eq.~(\ref{eq:species}), are solved for a reduced set of species, while the chemical source terms, $\widetilde{\dot{\omega}}_n$, are obtained from the $\epsilon$--$Z$ flamelet tables defined by Eq.~(\ref{eq:mean_tables_eps}). When $S^*$ exceeds the local flammability limit, the tabulated source terms are set to zero, $\widetilde{\dot{\omega}}_n=\widetilde{\dot{Q}}=0$, representing a locally non-reacting state. Product species formed upstream are nevertheless convected and diffused by the resolved transport equations. In a way, this may be interpreted as a multi-progress variable approach.

To reduce cost, only the dominant species are explicitly transported. For a detailed mechanism containing $M$ species, transport equations are solved for $N-1$ major species and one lumped residual species, where $N<M$. The lumped species is defined by
\begin{equation}\label{eq:gamma}
\xi = \sum_{n=1}^{N-1} \widetilde{Y}_n,
\end{equation}
the remaining fraction is represented by a composite lumped species,  
\begin{equation}\label{eq:prod}
\widetilde{Y}_{N} = 1 - \xi,
\end{equation}
with a source term ensuring mass conservation,  
\begin{equation}
\widetilde{\dot{\omega}}_{N} = -\sum_{n=1}^{N-1} \widetilde{\dot{\omega}}_n \:,
\end{equation}
\noindent so that $\sum_{n=1}^{N}\widetilde{\dot{\omega}}_n=0$.

In the present study, the detailed mechanism contains $M=13$ species. The explicitly transported set consists of $N-1=6$ major species: $\mathrm{O_2}$, $\mathrm{H_2O}$, $\mathrm{CH_4}$, $\mathrm{CO}$, $\mathrm{CO_2}$, and $\mathrm{N_2}$. The remaining species, $\mathrm{H_2}$, $\mathrm{H}$, $\mathrm{O}$, $\mathrm{OH}$, $\mathrm{HO_2}$, $\mathrm{CH_3}$, and $\mathrm{CH_2O}$, are represented by the lumped residual species $\widetilde{Y}_{N=7}$. This reduced set captures at least $95\%$ of the total mass fraction throughout the flamelet solution space (i.e., $\xi>0.95$), while avoiding resolved transport of the full mechanism and reducing the stiffness associated with short-lived radicals. The transported species mass fractions are used only in the evaluation of thermodynamic and transport properties, such as $C_{p,n}$, $R$, and $\mu$. They do not enter the evaluation of the chemical source terms, which are obtained directly from the flamelet tables. Consequently, the precise selection of species included in the explicitly transported set is not expected to significantly affect the combustion process, provided that $\xi$ remains close to unity.

\subsubsection{Treatment of Off-Manifold States}\label{sec:off_manifold}
\begin{figure}
    \centering
    \includegraphics[width=1.0\linewidth]{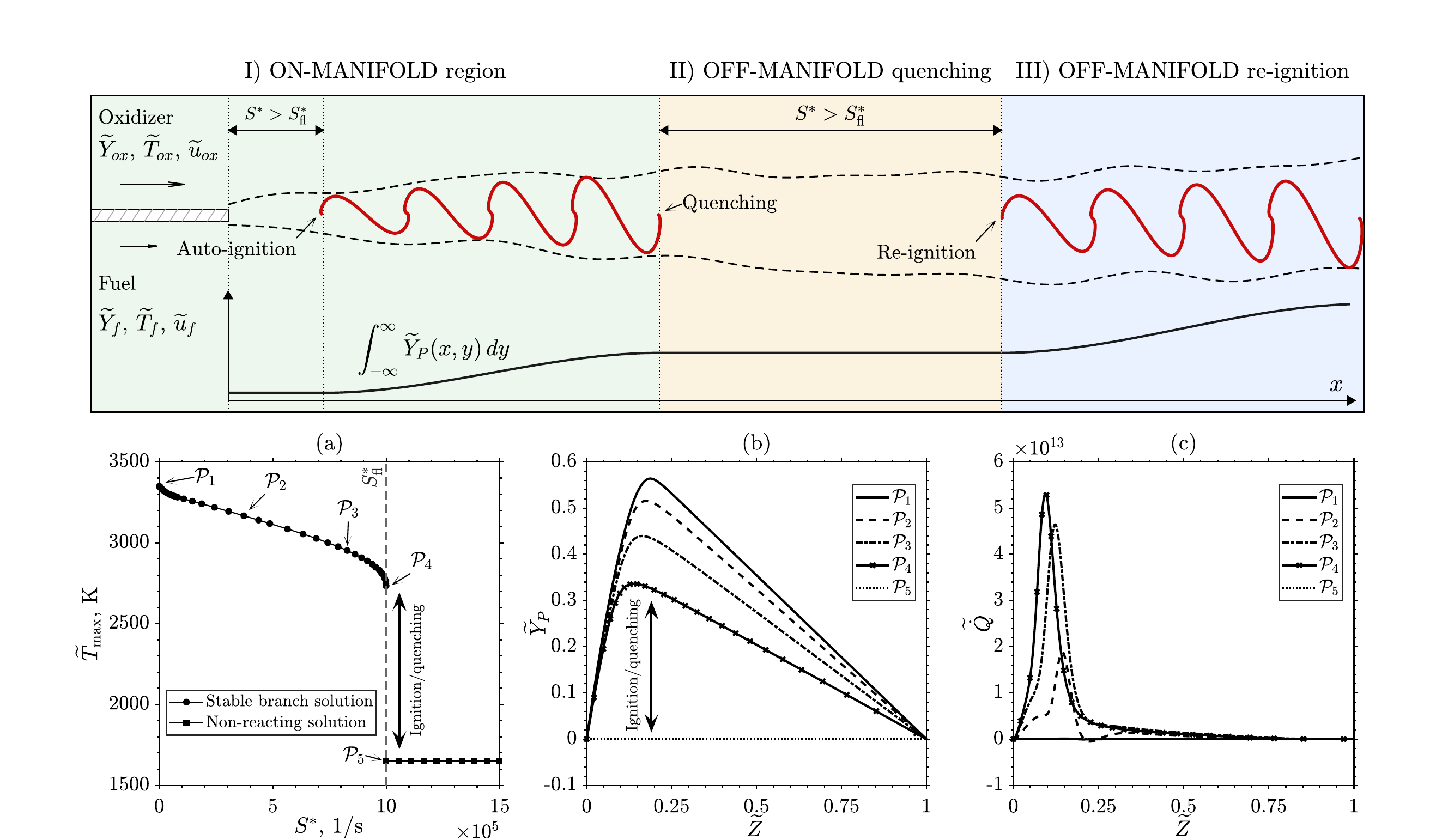}
    \caption{Schematic of on- and off- manifold states.}
    \label{fig:off-manifold}
\end{figure}

The use of explicit species transport with tabulated flamelet source terms introduces a distinction between the transported composition and the composition implied by the flamelet manifold. For a given set of table inputs, the transported mass fractions, $\widetilde{Y}_{n}$, need not coincide with the corresponding tabulated values. As a result, the tabulated source terms may be inconsistent with the locally transported composition and, in limiting cases, may attempt to consume more reactant than is locally available or produce a nonphysical total mass fraction. This issue is not removed by reducing the timestep, since the tabulated source terms are not explicit functions of the transported species mass fractions.

This behavior is most clearly illustrated during strain-rate-induced quenching and re-ignition, as shown schematically in Fig.~\ref{fig:off-manifold} for a reacting mixing layer with varying flamelet strain rate magnitude in the flow direction. The top graphic shows heat release in red limited to regions where the local flamelet strain rate $S^*$ is lower than the flammability limit strain rate $S^*_{\mathrm{fl}}$. Figure ~\ref{fig:off-manifold}a) shows the available tabulated manifold states in terms of maximum temperature as a function of strain rate. Five points of interest ($\mathcal{P}_{i=1,5}$) on this curve are identified. Figures ~\ref{fig:off-manifold}b) and ~\ref{fig:off-manifold}c) show the corresponding product mass fraction and heat-release rates, respectively, in terms of mixture fraction of these points.

In the on-manifold region (region I), the local flame structure can be represented by the tabulated flamelet solutions. The non-reacting state, $\mathcal{P}_{5}$, may represent the mixture prior to auto-ignition, while points $\mathcal{P}_{1}$--$\mathcal{P}_{4}$ represent reacting states following ignition. When the imposed strain rate exceeds the flammability limit, $S^*>S^*_{\mathrm{fl}}$ (region II), the tabulated source terms are set to zero and the local state is treated as non-reacting. However, products formed upstream continue to convect and diffuse through the quenched region. The resulting transported composition contains a memory of upstream reaction and therefore does not, in general, correspond to either a reacting flamelet solution or a non-reacting diffusion layer from the original library. Note that a flamelet model based on mixture fraction and strain rate without explicitly solving for species transport in the flow-field (such as the SLFM) cannot represent this region as $\mathcal{P}_{5}$ does not contain products. If the strain rate later decreases below the flammability limit (region III), re-ignition may occur from this off-manifold composition. In an FPV formulation, both off-manifold regions are represented indirectly by the transport progress-variable equation, which constrains the recovered species mass fractions to lie on the FPV manifold. In an ideal situation, the equilibrium solution $\mathcal{P}_{5}$ gets selected throughout region II, as the volumetric heat-release rate goes to zero in this case, however this may not be guaranteed as there is no mechanism to enforce this selection based on strain rate.  

Thus, the relevant issue is not merely a numerical inconsistency between transported and tabulated mass fractions, but rather that a flamelet library generated from fixed inlet compositions does not span all physically realizable transported states encountered after quenching and re-ignition. Expanding the flamelet library to include additional composition and temperature dimensions would formally address this limitation and has been explored by Hellwig~\cite{hellwig_three-dimensional_2026}. However, such an approach substantially increases computational cost, memory requirements, and model complexity, since additional state dimensions are needed to represent variations in species composition and temperature. For detailed hydrocarbon mechanisms, the resulting table dimensionality can quickly become prohibitive, motivating alternatives such as neural-network-based manifold representations. Instead, the present model retains a single composition/temperature flamelet library and treats the tabulated source terms as a nominal chemistry closure. When the transported state departs from the manifold, these source terms are corrected to enforce local species availability and mass conservation while preserving elemental consistency during the reaction update.

Here this correction is realized by defining a single local scaling parameter, $\alpha$, which is applied uniformly to all tabulated species production rates, 
\begin{equation} \widetilde{\dot{\omega}}_{n,\mathrm{corr}} = \alpha \widetilde{\dot{\omega}}_n, \qquad n = 1,2,\ldots,N . \label{eq:omega_alpha_corr} \end{equation} 
The same scaling is also applied to the tabulated heat-release rate, 
\begin{equation} \widetilde{\dot{Q}}_{\mathrm{corr}} = \alpha \widetilde{\dot{Q}} . \label{eq:q_alpha_corr} \end{equation} 
The parameter $\alpha$ is determined from the local fuel and oxidizer availability factors,
\begin{equation}
\alpha = \alpha_\mathrm{f} \alpha_{\mathrm{ox}},
\label{eq:alpha_product}
\end{equation}
where, for each reactant $r\in\{\mathrm{f},\mathrm{ox}\}$,
\begin{equation}
\alpha_r =
\begin{cases}
\displaystyle
\min\left(
\frac{\widetilde{Y}_r}{\widetilde{Y}_{r,\mathrm{TAB}}},
1
\right),
& \widetilde{\dot{\omega}}_r < 0, \\[2ex]
1,
& \widetilde{\dot{\omega}}_r \geq 0 .
\end{cases}
\label{eq:alpha_r}
\end{equation}
Here, $\widetilde{Y}_\mathrm{f}$ and $\widetilde{Y}_{\mathrm{ox}}$ are the transported mean fuel and oxidizer mass fractions determined from Eq.~(\ref{eq:species}), whereas $\widetilde{Y}_{\mathrm{f},\mathrm{TAB}}$ and $\widetilde{Y}_{\mathrm{ox},\mathrm{TAB}}$ denote the corresponding tabulated mean mass fractions for the local values of $\widetilde{Z}$, $\widetilde{Z''^2}$, $S^*$, and $\bar{p}$ in a given computational cell. In this study $\mathrm{f=CH_4}$ and $\mathrm{ox=O_2}$. The correction therefore leaves the tabulated source terms unchanged when the transported reactant abundances are at least as large as those implied by the flamelet table, or when the corresponding reactant is not being consumed. The full source-term vector is reduced only when fuel or oxidizer consumption would otherwise be inconsistent with the locally available transported mass fraction.

This is justified by the elementary-reaction structure of the dominant reactant-consumption terms. For an elementary reaction in which a reactant \(X\) appears with unit stoichiometric coefficient,
\begin{equation}
A+X \rightleftharpoons C+D,
\end{equation}
the forward progress rate is
\begin{equation}
r_f = k_f [A][X],
\end{equation}
where \(k_f\) is the forward reaction-rate constant and \([\cdot]\) denotes molar concentration. The destruction contribution to the source term of \(X\) is therefore
\begin{equation}
\dot{\omega}_X^- = -W_X r_f
                  = -W_X k_f [A][X],
\end{equation}
where the superscript "$-$" indicates destruction. The destruction rate is thus first order in \(X\), implying that a reduction in the local reactant concentration leads to an approximately proportional reduction in its consumption rate. Consequently, if the transported reactant abundance is smaller than that implied by the flamelet table, the corresponding reaction rate may be approximated as scaling linearly with the ratio \(Y_X/Y_{X,\mathrm{TAB}}\).

For detailed hydrocarbon oxidation mechanisms, fuel and oxidizer consumption is generally dominated by bimolecular elementary reactions in which the consumed reactant appears once on the reactant side. The product form in Eq.~(\ref{eq:alpha_product}) therefore provides a physically motivated local availability correction that accounts for reactant depletion while preserving the tabulated chemistry structure. The correction should be interpreted as a bounded reactant-availability limiter rather than a reconstruction of the full detailed-chemistry source term. This approach is based on the work of Ihme et al.~\cite{ihme_modeling_2008}, who applied a similar correction to the nitric oxide source term in a flamelet/progress-variable formulation. In that study, the NO destruction rate was scaled according to the transported NO mass fraction to account for departures from the tabulated state. Comparisons with experimental measurements showed improved agreement when NO was transported explicitly with the corrected source term, relative to directly prescribing the NO mass fraction from the flamelet tables. 

\subsection{Combustion Models Summary}

Table~\ref{tab:summary} summarizes the OSK, SLFM, FPV, and $\epsilon$--$Z$ combustion-model formulations, together with their associated transport equations for RANS computations. The SLFM is included for comparison, although it is not used in the present work. The number of species, $N$, and the corresponding number of transport equations are reported for the reaction mechanism considered here; these values would naturally change for different chemical mechanisms and applications.

Flamelet-based models offer a clear advantage by incorporating detailed chemical kinetics without requiring a proportional increase in the number of transported equations. For example, direct CFD transport of the same 13-species mechanism used in the flamelet libraries would require 20 transport equations, compared with 10--14 equations for the flamelet-based approaches considered here. This advantage becomes more significant as the size of the chemical mechanism increases.

The proposed $\epsilon$--$Z$ formulation, however, introduces additional cost relative to FPV. In FPV, the number of transported thermochemical variables, $\widetilde{Z}$, $\widetilde{Z''^2}$, and $\widetilde{C}$, is independent of the chemical-mechanism size. In contrast, the $\epsilon$--$Z$ model solves a subset of species transport equations, and the number of transported species may increase with mechanism size in order to keep $\gamma$, defined in Eq.~(\ref{eq:gamma}), close to unity. For example, Walsh et al.~\cite{walsh_flamelet_2026} applied the $\epsilon$--$Z$ model to JP-5/air combustion in a turbine passage using a 118-species mechanism while transporting a subset of 14 species. Although this still represents a substantial reduction relative to full species transport, the $\epsilon$--$Z$ model remains more expensive than FPV. In the present two-dimensional RANS cases, this added cost is not significant; however, its scaling for three-dimensional LES computations remains to be quantified and may require a dedicated cost study.
\newpage
\begin{landscape}
\begin{table}[p]
\centering
\caption{Summary of combustion model formulations and associated transport equations.}
\label{tab:summary}
\renewcommand{\arraystretch}{1.25}
\small

\begin{tabularx}{\linewidth}{|c|>{\centering\arraybackslash}p{0.08\linewidth}|Y|Y|Y|Y|Y|Y|}
\hline
&
\textbf{Species, $\mathbf{N}$} &
\textbf{Species transport} &
\textbf{Chemical source terms} &
\textbf{Inputs to flamelet tables} &
\textbf{Scalars $\boldsymbol{\psi}_j$ brought to the computation} &
\textbf{Total number of transport equations} &
\textbf{Treatment of off-manifold flamelet states} \\
\hline

\textbf{OSK} &
5 &
Explicitly, through species transport equations &
From a single global reaction based on the mean species concentrations &
N/A &
N/A &
12 (5 RANS, 2 $k-\omega$ and 5 species) &
N/A
\\
\hline

\textbf{SLFM} &
$13$* &
Implicitly, through the $\widetilde{Z}$ equation &
From 13 species and 32 reactions (sub-grid chemistry) &
$\widetilde{Z}$, $\widetilde{Z''^2}$, $\chi_{st}(\widetilde{\chi})$ and $\bar{p}$ &
$\widetilde{Y}_{n}$ and $\widetilde{\dot{Q}}$ &
9 (5 RANS, 2 $k-\omega$, $\widetilde{Z}$ and $\widetilde{Z''^2}$) &
Forces a tabulated manifold state**** 
\\
\hline

\textbf{FPV} &
$13$* &
Implicitly, through the $\widetilde{Z}$ and $\widetilde{C}$ equations &
From 13 species and 32 reactions (sub-grid chemistry) &
$\widetilde{Z}$, $\widetilde{Z''^2}$, $\widetilde{C}$ and $\bar{p}$ &
$\widetilde{Y}_{n}$, $\widetilde{C}_{\mathrm{tab}}$*** and $\widetilde{\dot{Q}}$ &
10 (5 RANS, 2 $k-\omega$, $\widetilde{Z}$, $\widetilde{Z''^2}$ and $\widetilde{C}$) &
Forces a tabulated manifold state through advection and diffusion of $\widetilde{C}$
\\
\hline

\textbf{$\boldsymbol{\epsilon}$--$\boldsymbol{Z}$} &
$7$** &
Explicitly, through species transport equations &
From 13 species and 32 reactions (sub-grid chemistry) &
$\widetilde{Z}$, $\widetilde{Z''^2}$, $S^*(\epsilon)$ and $\bar{p}$ &
$\widetilde{\dot{\omega}}_{n}$ and $\widetilde{\dot{Q}}$ &
14 (5 RANS, 2 $k-\omega$, $\widetilde{Z}$, $\widetilde{Z''^2}$ and 7 species) &
Scales the chemical rates for a given combination of $\widetilde{Z},S^*,\widetilde{Z''^2},\bar{p}$ based on reactant availability
\\
\hline
\end{tabularx}

\vspace{0.5em}
\begin{minipage}{\linewidth}
\footnotesize
* Corresponding to the number of species, $M$, considered in the reaction mechanism used to solve the flamelet equations.\\
** Corresponding to the subset of the $M$ species that are tracked during the computation.\\
*** $\widetilde{C}_{\mathrm{tab}}$ is the tabular progress variable used for the determination of $\lambda$.\\
**** With discontinuities in $\widetilde{Y}$ between regions where $\chi_{\mathrm{st}}(\widetilde{\chi}) > \chi_{\mathrm{st},\max}$.
\end{minipage}
\end{table}
\end{landscape}

%% file: sections/results.tex
\section{Computational Results and Discussions}\label{sec:results}
This section presents the results of the reacting flow within the nozzle, obtained using the OSK, FPV and $\epsilon$--$Z$ combustion models. 
\subsection{Mean Flow Structure}
\begin{figure}[h]
\centering
\begin{subfigure}{.45\textwidth}
  \centering
  \includegraphics[width=1.0\textwidth]{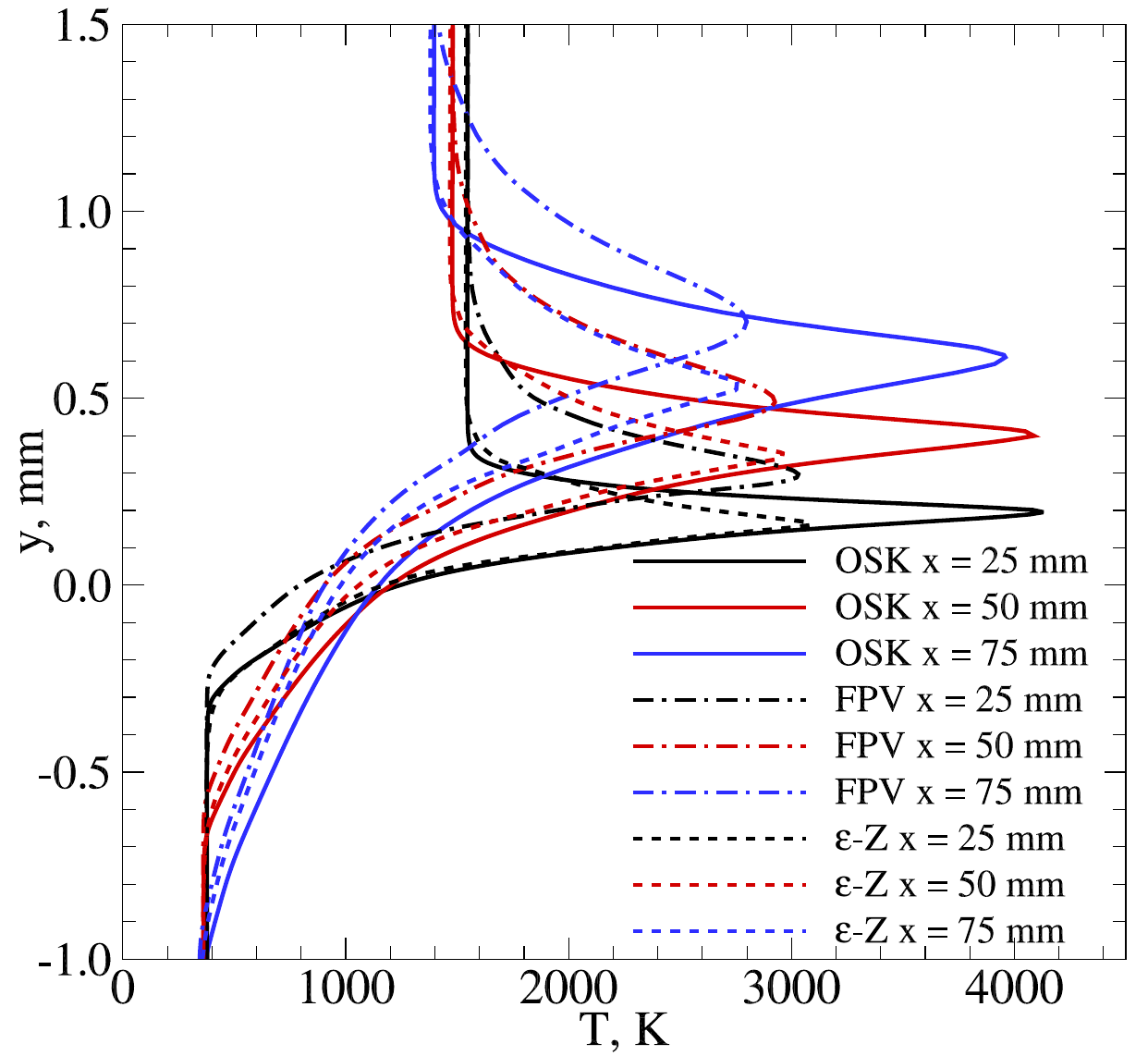}
  \caption{Temperature.}
  \label{fig:temp_lines}
\end{subfigure}%
\begin{subfigure}{.45\textwidth}
  \centering
  \includegraphics[width=1.0\textwidth]{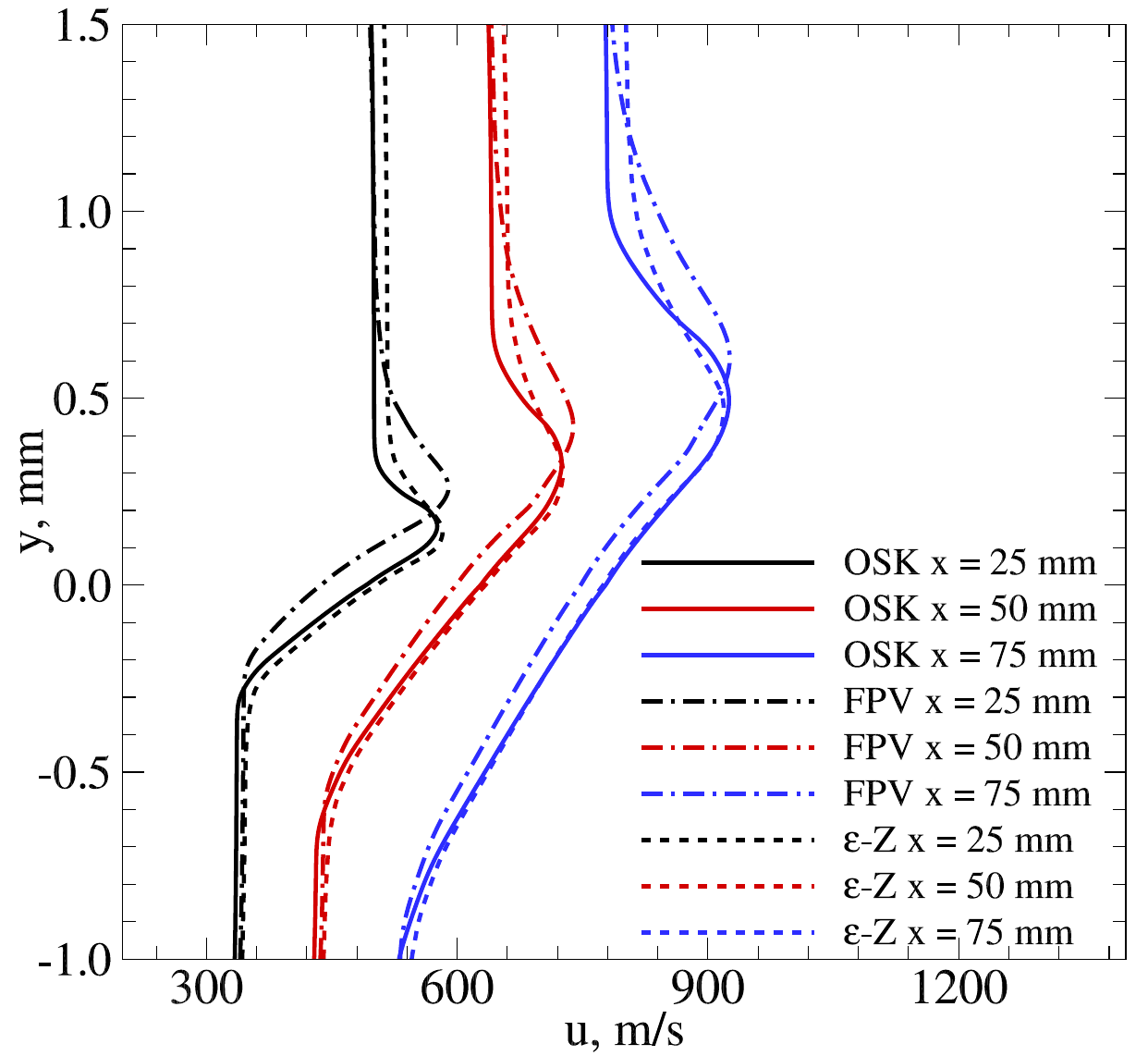}
  \caption{Streamwise velocity.}
  \label{fig:u_lines}
\end{subfigure}
\caption{Profiles of temperature and velocity at three different streamwise locations as predicted using the OSK, FPV and $\epsilon$--$Z$ combustion models.}
\label{fig:mixing_layers}
\end{figure}

At the trailing edge of the splitter plate ($x$ = 3 mm), the hot oxidizer stream above begins to mix with the colder fuel vapor stream below, forming thermal and velocity mixing layers that extend downstream. Ignition occurs where the two streams merge, establishing a non-premixed flame within the mixing layer, with the reaction zone located near the stoichiometric mixture composition. Figure \ref{fig:mixing_layers} presents the temperature and streamwise velocity profiles at three downstream positions. The streamwise velocity attains its maximum value in the high-temperature region owing to the local decrease in density. As the flow progresses, the mixing layer thickens and the freestream velocity increases due to flow acceleration under the large favorable pressure gradient. The air-side mixing layer is noticeably thicker than the fuel-side layer, which can be attributed to the lower density and higher molecular viscosity of the hot oxidizer stream, resulting in a smaller local Reynolds number despite its higher velocity. The thermal and velocity layer thicknesses remain approximately similar, consistent with the use of identical turbulent Schmidt and Prandtl numbers in the computations.

Both flamelet-based models predict substantially lower flame temperatures than the OSK model, with peak temperatures reduced by approximately $1000~\mathrm{K}$. This difference is expected because the flamelet formulations include detailed finite-rate chemistry, including dissociation and other non-equilibrium effects that reduce the attainable flame temperature. A similar trend was reported by Walsh et al.~\cite{walsh_turbulent_2025}. The FPV results also exhibit substantially thicker thermal and velocity mixing layers than the other two models as well as a vertical shift towards the air side (top). This behavior is associated with the broader reaction zone predicted by FPV, where the spatial extent of the source terms is governed not only by the mixture-fraction field but also by the distribution of the flamelet parameter $\lambda$. This point is examined in more detail below. The $\epsilon$--$Z$ flamelet model produces thermal and velocity layers that closely match the OSK results for layer-width magnitudes and layer locations. A more detailed comparison of these features is presented in the following discussion.

\begin{figure}
    \centering
    \includegraphics[width=1.0\linewidth]{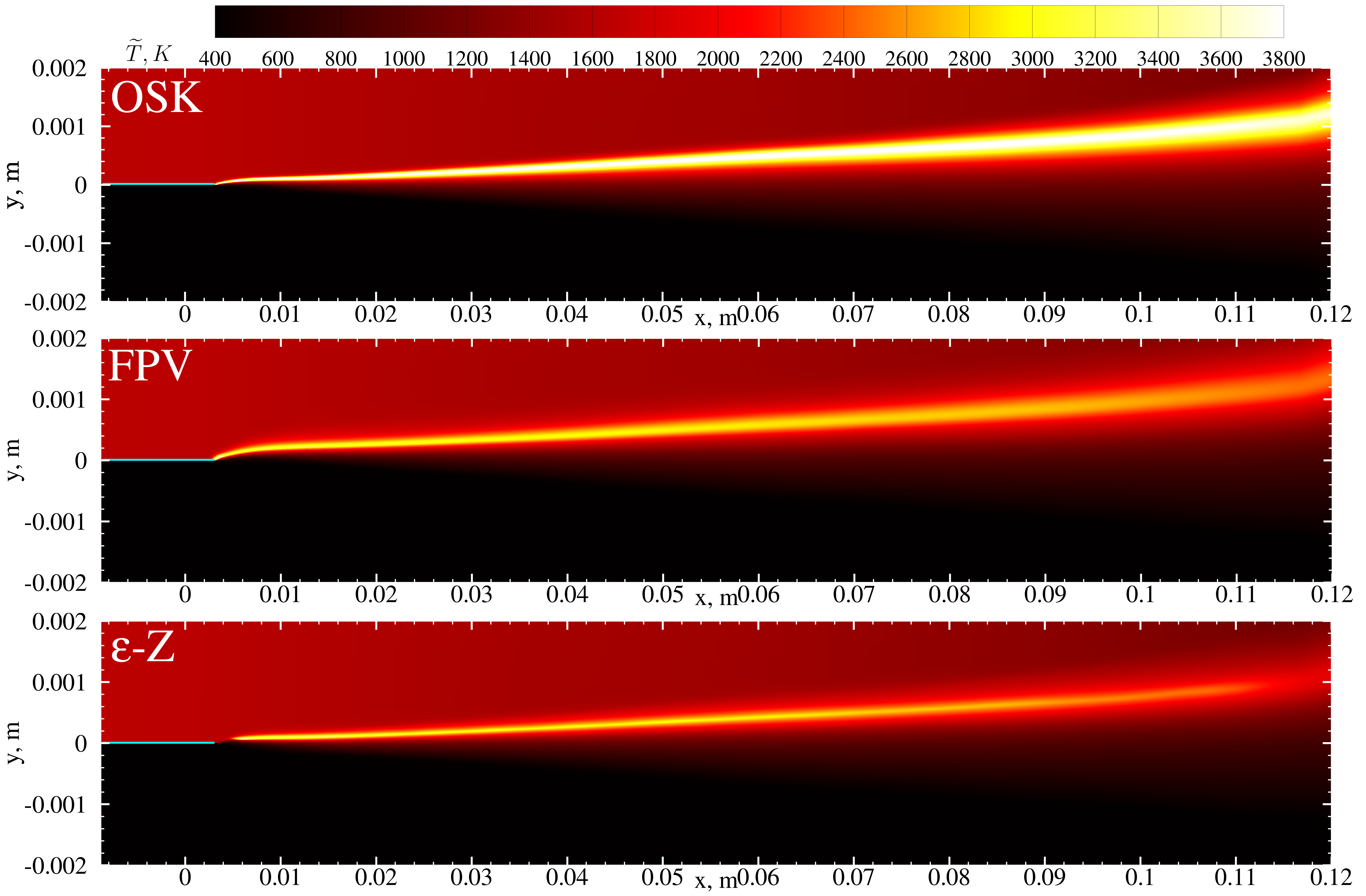}
    \caption{Temperature contours for OSK (top), FPV (center) end $\boldsymbol{\epsilon}$-based (bottom) combustion models.}
    \label{fig:tempcont}
\end{figure}

Figure~\ref{fig:tempcont} presents the temperature fields predicted by the three combustion models. Here, the splitter plate is identified as the blue horizontal lines near the inlets. The transverse direction has been magnified to highlight the flame structure in greater detail. Overall, the three models exhibit qualitatively similar behavior, aside from the lower peak temperatures predicted by the flamelet models, as discussed previously. A key difference among the models lies in the flame standoff distance. The OSK and FPV models predicts zero standoff, with ignition occurring immediately at the trailing edge of the splitter plate.  The $\epsilon$--$Z$ flamelet model, in contrast, predicts a distinct flame standoff of about 2 mm.

In the latter case, the flame standoff arises because the local turbulence kinetic energy dissipation rate, $\epsilon$, is large in the vicinity of the splitter plate, where strong shear develops due to the no-slip boundary condition at the wall and the large velocity gradients between the two streams. Immediately downstream of the plate, the corresponding flamelet strain rate, $S^*$, exceeds the flamelet flammability limit, $S^*_{\mathrm{fl}}$, leading to local extinction and delaying ignition until the flow reaches a region where $\epsilon$ reduces and $S^*<S^*_{\mathrm{fl}}$. By contrast, the predicted flame standoff (or for this case the lack thereof) by the FPV model is governed by the growth rate of the transported progress variable and is therefore not explicitly constrained by the local strain rate. The growth rate of the progress variable depends on the chosen definition of $C$ (see Eq.~(\ref{cequation})), since different definitions yield different source terms, $\widetilde{\dot{\omega}}_C$, and consequently different rates of progress-variable evolution. As a result, the predicted flame standoff distance in the FPV formulation is sensitive to the particular choice of progress variable, and an alternative definition of $C$ could produce a distinct flame anchoring location.

Additionally, the $\epsilon$--$Z$ model exhibits a pronounced temperature drop near x = 110 mm, where the static pressure is approximately 9 bar. This drop corresponds to local flamelet quenching, which arises from the decrease in the flammability limit with pressure, $S^*_{\mathrm{fl}}(\bar{p})$ (as indicated in Fig. \ref{fig:tmax_sstar}). Both this downstream quenching and the previously observed flame standoff reflect a direct response of the flamelet to the strain field predicted by the mean flow, as intended in the $\epsilon$--$Z$ formulation. Neither the OSK nor the FPV model can capture this behavior, as both lack the required mechanism, as discussed below.

\subsection{$\boldsymbol{\epsilon}$--$\mathbf{Z}$ Flamelet Strain-Rate Coupling}
\begin{figure}
    \centering
    \includegraphics[width=1.0\linewidth]{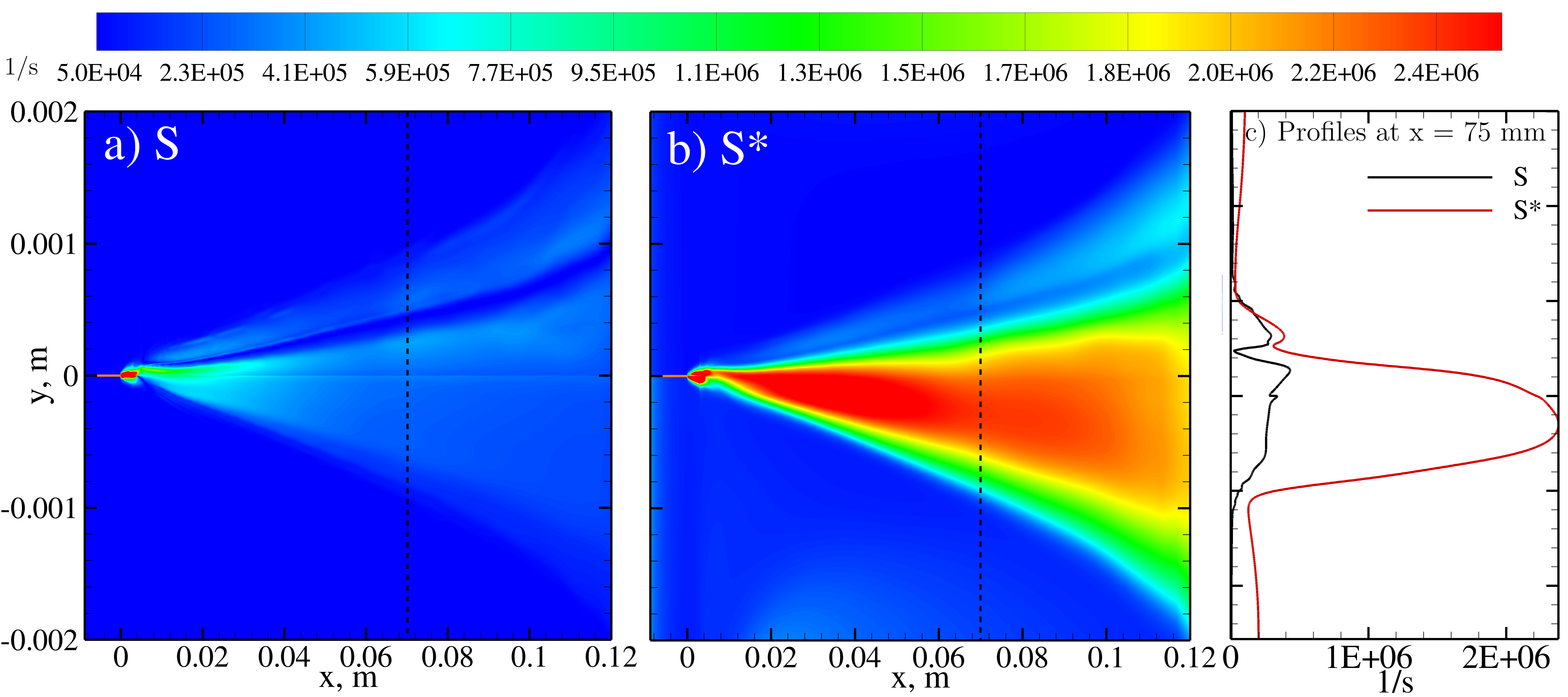}
    \caption{Left a): contours of the mean strain-rate magnitude, ${S}$. Center b): contours of the flamelet inflow strain rate ${S^*}$. Right c): profiles of ${S}$ and ${S^*}$ at the streamwise location of x = 75 mm (represented by the black vertical lines in a) and b)).}
    \label{fig:sr_combined}
\end{figure}

Figure~\ref{fig:sr_combined}a presents contours of the mean strain-rate magnitude, $S$, for the $\epsilon$--$Z$ computation. $S$ here follows the same definitions as in Eq. (\ref{eq:turbulent_Pk}). The $S$ fields for the OSK and FPV results exhibit similar spatial distributions. The splitter plate (represented by the orange horizontal line) extends from the inlet to $x$ = 3 mm, but the no-slip boundary condition is applied only between $x$ = 0 mm and $x$ = 3 mm. Within this region, the strong velocity gradients induced by the wall generate the highest strain rates, which appear as the red zone in the contour plot. Downstream of the splitter plate, two high-strain regions develop on either side of the flame, separated by a low-strain trough located at the flame centerline. These regions of elevated strain correspond to the strong velocity gradients that form on both sides of the reaction zone, while the local minimum coincides with the flame core, where the streamwise velocity reaches its maxima and strain-rate production is minimized. A profile of $S$ evaluated at $x$ = 75 mm is shown in Fig. \ref{fig:sr_combined}c.

Figure~\ref{fig:sr_combined}b shows the corresponding contours of the flamelet strain rate, $S^*$, computed using $\epsilon$ according to Eq.~(\ref{eq:sstar-eps}). Its profile at $x$ = 75 mm is shown in Fig. \ref{fig:sr_combined}c. A comparison with the mean strain rate magnitude $S$ reveals that $S^*$ exhibits a similar spatial structure, featuring two high-strain regions flanking the flame and a trough at the flame location. With this spatial distribution, the subgrid flamelet dynamics respond consistently to the mean strain field. The magnitudes of $S^*$ are systematically higher than those of $S$, reflecting the increase in strain intensity at smaller turbulent scales, consistent with the energy cascade behavior of turbulence. This behavior is made possible by employing $\epsilon$ as the flamelet tracking variable, which establishes a direct link between the turbulence dissipation and the small-scale flamelet strain rate.

\begin{figure}
\centering
\begin{subfigure}{.5\textwidth}
  \centering
  \includegraphics[width=1.0\textwidth]{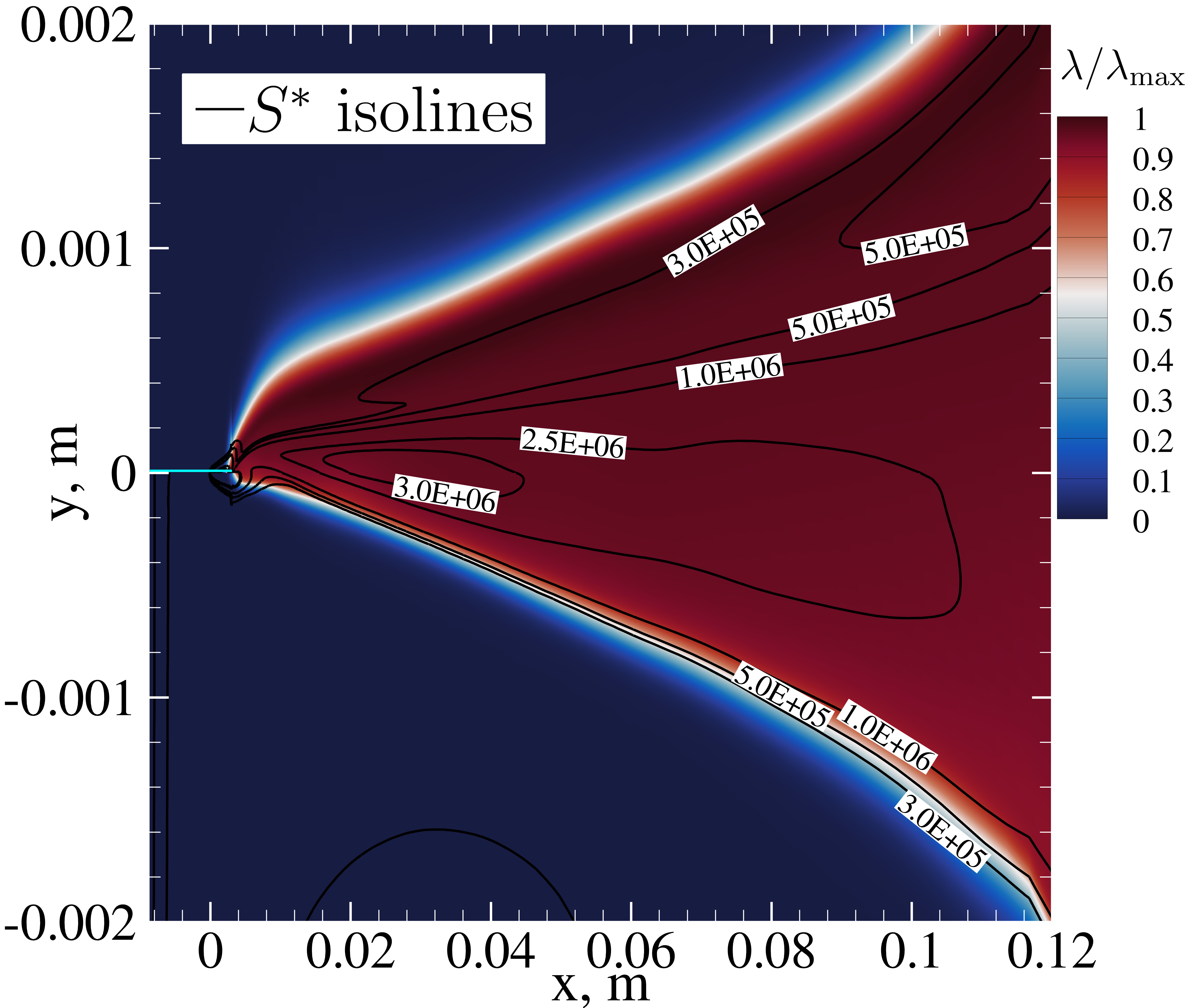}
  \caption{Contour of the normalized flamelet parameter ${\lambda/\lambda_{{max}}}$.}
  \label{fig:fpv_lambda}
\end{subfigure}%
\begin{subfigure}{.5\textwidth}
  \centering
  \includegraphics[width=0.92\textwidth]{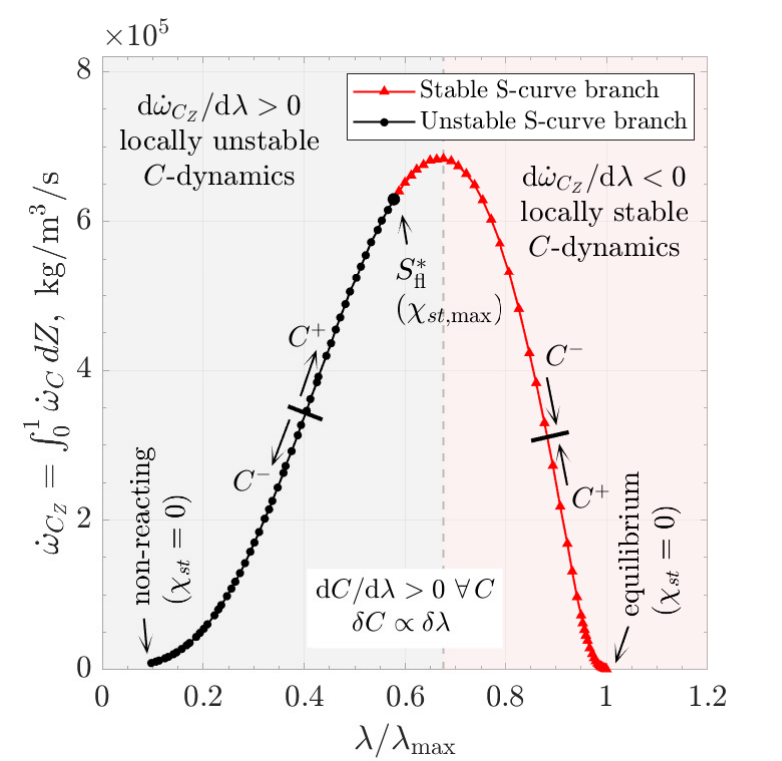}
  \caption{Integrated progress-variable source term, $\dot{{\omega}}_{{C_Z}}({\lambda})$.}
  \label{fig:c_stability}
\end{subfigure}
\caption{(a) Spatial distribution of the normalized flamelet parameter and (b) Mixture-fraction-integrated progress-variable source term.}
\label{fig:C_stability_and_lambda}
\end{figure}

\subsection{FPV Flamelet-State Selection and Strain-Rate Inconsistency}
In the FPV framework, the transported mean progress variable, $\widetilde{C}$, is typically used to document the coupling between the flow solver and the flamelet library and is therefore the quantity most commonly reported. However, for a given combination of $\widetilde{Z}$, $\widetilde{Z''^2}$, and $\bar{p}$, the corresponding monotonic relation $\widetilde{C}(\lambda)$ must be inverted to recover the flamelet parameter $\lambda$, which determines the location of the local flamelet solution on the ``s-shaped" curve and, consequently, its associated flamelet strain rate.

Figure~\ref{fig:fpv_lambda} shows the spatial distribution of the normalized flamelet parameter, $\lambda/\lambda_{\max}$, obtained from the inversion of $\widetilde{C}(\lambda)$. Isolines of the corresponding flamelet strain rate, $S^*$, are overlaid, and the splitter plate is indicated by the blue horizontal line near the inlet. The distribution should be interpreted in conjunction with Fig.~\ref{fig:qmapping}, which relates $\lambda/\lambda_{\max}$ to the corresponding flamelet $\chi_{st}$ or strain rate $S^*$. The colormap is chosen such that white regions correspond to the value of $\lambda/\lambda_{\max}$ associated with the flamelet flammability limit, $S^*_{\mathrm{fl}}$. Along lines of constant $x$, moving inward from the far field on either the oxidizer or fuel side, $\widetilde{C}=0$ and the inversion therefore gives $\lambda/\lambda_{\max}=0$, corresponding to the non-reacting flamelet solution. Near the edge of the mixing layer, approximately indicated by the $3\times10^5~\mathrm{s^{-1}}$ isoline of $S^*$, $\lambda/\lambda_{\max}$ rapidly approaches unity from both sides of the mixing layer and remains close to unity throughout the mixing layer. In the streamwise direction, $\lambda/\lambda_{\max}$ reaches unity immediately downstream of the splitter plate and remains near this value farther downstream.

This distribution implies that the FPV model samples maximum-strain-rate flamelet solutions, indicated by the white regions, near the edges of the mixing layer, where the local strain rate is small. Conversely, within the mixing layer, where the inferred strain rates are largest, the FPV model predominantly samples flamelet states located near the equilibrium branch of the S-curve, corresponding to very low flamelet strain rates. This behavior persists despite substantial spatial variation in the inferred flamelet strain rate, with contours ranging from approximately $3\times10^6~\mathrm{s^{-1}}$ near the splitter plate to $3\times10^5~\mathrm{s^{-1}}$ farther downstream.

This behavior can be interpreted with Fig. \ref{fig:c_stability} using the mixture-fraction-integrated progress-variable source term,
\begin{equation}
\dot{\omega}_{C_Z}(\lambda)
=
\int_0^1 \dot{\omega}_C(Z,\lambda)\rm{d}Z .
\end{equation}
Here, this quantity is shown for a background pressure of $30~\mathrm{bar}$ and in the limit $\widetilde{Z''^2}\rightarrow 0$; however, the analysis is unchanged for other combinations of pressure and mixture-fraction variance. The progress-variable mapping is monotonic in $\lambda$ such that $d\widetilde{C}/{d\lambda} > 0 \; \forall\widetilde{C}$. Therefore, a perturbation in the transported progress variable corresponds directly to a perturbation in the flamelet coordinate, $\delta \widetilde{C}\propto\delta\lambda$, and the sign of $d\dot{\omega}_{C_Z}/d\widetilde{C}$ is determined by the sign of $d\dot{\omega}_{C_Z}/d\lambda$. On the rising portion of $\dot{\omega}_{C_Z}(\lambda)$, where $d\dot{\omega}_{C_Z}/d\lambda>0$, perturbations in $\widetilde{C}$ are amplified by the homogeneous source dynamics. On the descending portion, where $d\dot{\omega}_{C_Z}/d\lambda<0$, a positive perturbation in $\widetilde{C}$ produces a smaller source term and a negative perturbation produces a larger source term, so perturbations are locally damped. The high-$\lambda$ portion of the manifold therefore behaves as a locally attracting region of the homogeneous progress-variable dynamics.

This source-term structure provides a plausible explanation for the observed collapse of $\lambda/\lambda_{\max}$ toward unity throughout the interior of the mixing layer. Once the solution passes the flammability-limit region, the integrated progress-variable source remains positive and continues to drive $\widetilde{C}$, and hence $\lambda$, toward the equilibrium state. At the same time, the negative slope of $\dot{\omega}_{C_Z}(\lambda)$ on the high-$\lambda$ side damps local perturbations in the homogeneous source dynamics. As a result, the transported FPV solution preferentially relaxes toward the equilibrium solution over much of the reacting layer, even though the inferred $S^*$ field exhibits substantial spatial variation. This interpretation should not be read as a global stability statement for the full transported progress-variable equation. In the CFD solution, advection and diffusion can move the local value of $\widetilde{C}$ away from the chemically attracting region, and local changes in $\widetilde{Z}$, $\widetilde{Z''^2}$, and $\bar{p}$ also modify the relevant flamelet mapping. Nevertheless, the predominance of $\lambda/\lambda_{\max}\approx 1$ in the reacting interior indicates that the source-term dynamics strongly favor relaxation toward the equilibrium side of the FPV manifold once the solution has ignited.

The result highlights a fundamental characteristic of the FPV formulation. Although the local flamelet strain rate varies significantly throughout the flowfield, the transported progress variable evolves toward values associated with $\lambda/\lambda_{\max}\approx1$, causing the model to remain concentrated near equilibrium flamelet states. Consequently, the flamelet state selected by the FPV model is only weakly connected to the local strain-rate environment, in contrast to the present $\epsilon$--$Z$ formulation, where the flamelet strain rate is prescribed directly through the turbulence field.

\begin{figure}
\centering
\centering
\includegraphics[width=1.0\textwidth]{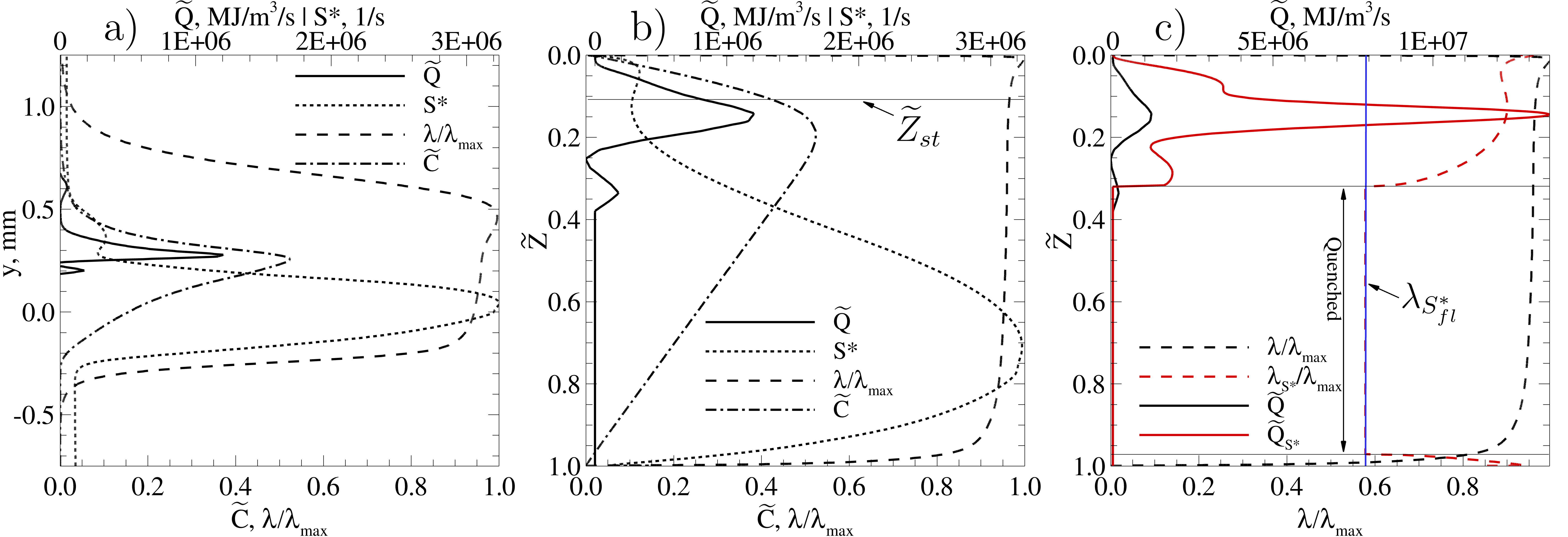}
\caption{Quantities of interest for the FPV results at the streamwise location x = 25 mm.}
\label{fig:fpv_lines}
\end{figure}

Figure~\ref{fig:fpv_lines} examines the FPV progress-variable coupling at the streamwise location $x=25$ mm. Panel a) shows the mean heat-release rate, $\widetilde{\dot{Q}}$, the strain rate inferred from the local turbulent dissipation, $S^*$ (with a spatial distribution that follows $S$), the normalized flamelet parameter, $\lambda/\lambda_{\max}$, and the transported mean progress variable, $\widetilde{C}$, as functions of the transverse coordinate $y$. Panel b) shows the same quantities plotted against the mean mixture fraction, $\widetilde{Z}$. The transported progress variable exhibits a localized, Gaussian-like distribution within the reacting layer, where its source term is active. Inverting the tabulated relation $\widetilde{C}(\lambda)$ gives values of $\lambda/\lambda_{\max}$ close to unity over much of the reacting region. These values correspond to flamelet solutions near the equilibrium, zero-strain limit, despite the fact that the local strain rate $S^*$ remains finite.

This behavior affects the spatial distribution of the FPV chemical source terms. As shown in Fig.~\ref{fig:qmapping}, the tabulated heat-release rate increases as $\lambda$ decreases from its maximum value. Therefore, in the FPV formulation, the spatial extent and location of $\widetilde{\dot{Q}}$ are governed not only by $\widetilde{Z}$, but also by the inferred value of $\lambda$. In the present case, the slight reduction of $\lambda$ toward larger $\widetilde{Z}$ enhances the tabulated source terms on the fuel side, resulting in a broader and shifted reaction zone away from the stoichiometric mixture fraction, $\widetilde{Z}_{st}=0.11$. This contributes to the broader reacting layer predicted by FPV compared with the OSK and $\epsilon$--$Z$ models, as quantified below.

Panel c) compares the FPV-coupled quantities with a post-processed strain-rate-based mapping. The black curves show the original FPV values of $\lambda/\lambda_{\max}$ and corresponding $\widetilde{\dot{Q}}$. The red curves show the analogous quantities, $\lambda_{S^*}/\lambda_{\max}$ and $\widetilde{\dot{Q}}_{S^*}$, obtained by using the local $S^*$ to select a flamelet state from the relation $\lambda(S^*)$ in Fig.~\ref{fig:qmapping}. The upper stable branch is used in this comparison, since $S^*$ alone does not uniquely determine the flamelet state on the S-curve. Unlike the FPV-inferred $\lambda$, the strain-rate-based value varies strongly across the mixing layer: it approaches unity near the non-reacting edges, decreases as $S^*$ increases, and reaches the flammability-limit value $\lambda_{S^*_{\mathrm{fl}}}$ where the imposed strain rate equals the maximum sustainable flamelet strain rate. For larger strain rates, the corresponding flamelet solution is quenched. The resulting $\widetilde{\dot{Q}}_{S^*}$ is significantly larger than the FPV value over much of the reacting region, because the FPV inversion places the local flamelet state close to equilibrium, where the volumetric chemical rates are smaller. This comparison should not be interpreted as a direct prediction that a strain-rate-based flamelet model would produce an order-of-magnitude larger heat release in the CFD solution, since the red curves are not coupled to the flow solver. Rather, the comparison illustrates the degree to which the FPV-inferred flamelet parameter differs from the value that would be implied by the local strain rate. It also shows that the FPV formulation can retain finite heat release in regions where the local $S^*$ exceeds the flamelet flammability limit, as evidenced by the $\widetilde{\dot{Q}}$ present in the quenched region, near $\widetilde{Z}=0.4$

Finally, this issue is not specific to the form of the energy equation. Many FPV formulations solve an energy equation written in terms of sensible plus formation enthalpy, for which the heat-release term does not appear explicitly. In that case, the heat release is represented implicitly through the tabulated species composition. However, if the flamelet state is still selected using $\widetilde{C}$, the retrieved composition remains inconsistent with the local strain-rate condition. Therefore, the same coupling issue persists regardless of whether heat release appears explicitly in the energy equation.

\subsection{Heat-Release-Rate Fields and Statistics}

\begin{figure}
\centering
\includegraphics[width=0.85\textwidth]{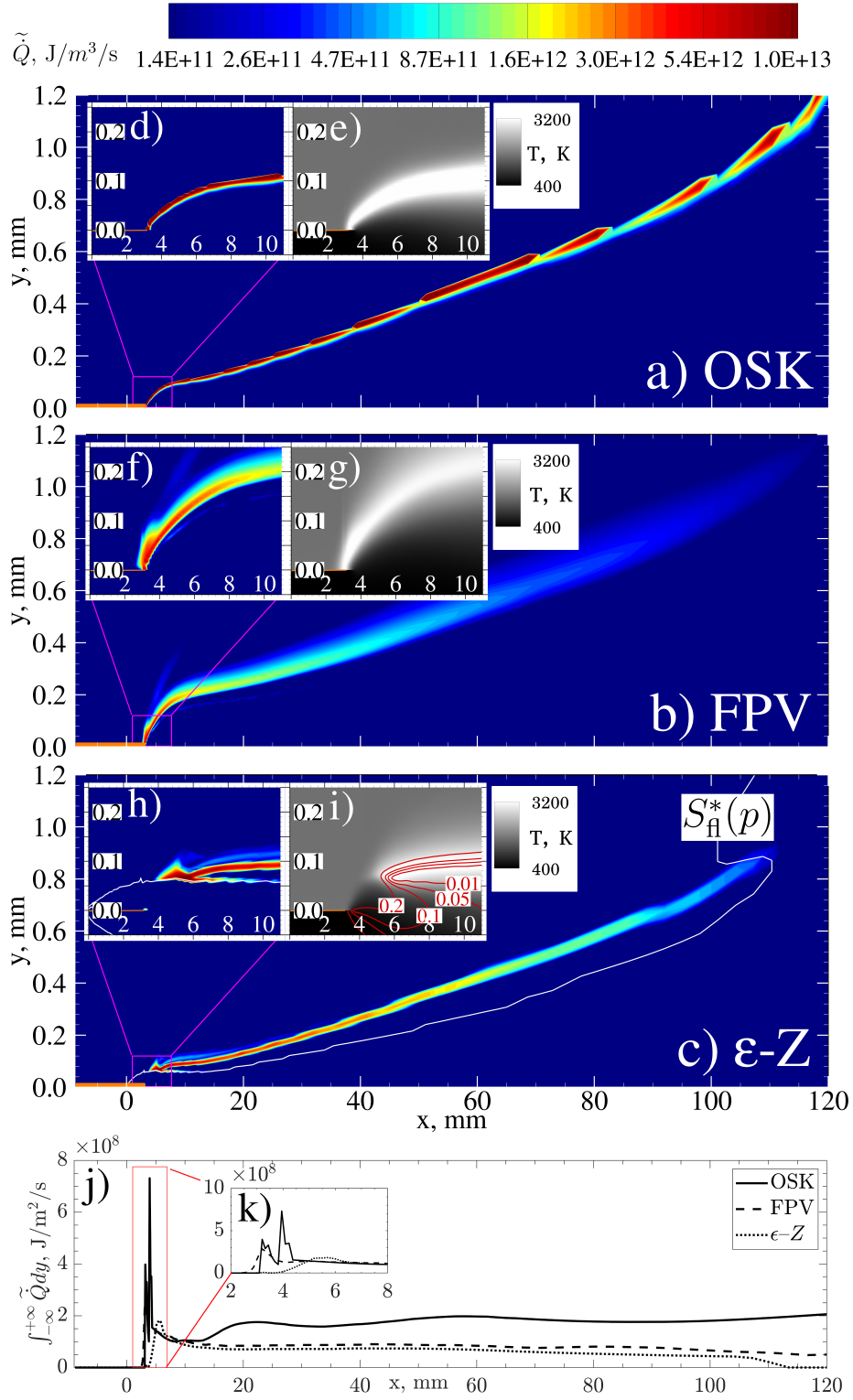}
\caption{
Heat-release rate, $\widetilde{\dot{{Q}}}$, predicted by the OSK, FPV, and ${\epsilon}$--${Z}$ combustion models.
}
\label{fig:heat_releases}
\end{figure}

Figure~\ref{fig:heat_releases} compares the heat-release-rate predictions by the three combustion models. The OSK, FPV, and $\epsilon$--$Z$ contours of heat-release rate are shown panels a), b) and c), respectively. The inset panels d), f), and h) provide close-up views of the splitter-plate region, where ignition first occurs, as indicated by the magenta boxes. Panels e), g), and i) show the corresponding temperature fields in the same region. The splitter plate is shown by the orange line in each panel. Finally, panel j) shows the heat-release rate integrated over the transverse direction, $y$, as a function of the streamwise direction, $x$, with a close-up near the ignition location in panel k).

The OSK formulation produces the largest heat-release rates, consistent with its higher predicted temperatures. In contrast, both flamelet-based models yield lower peak heat release because the tabulated chemistry includes detailed finite-rate effects, including dissociation and non-equilibrium behavior. However, the FPV model predicts substantially lower peak heat-release rates than the $\epsilon$--$Z$ model, despite both models using the same underlying flamelet solutions. This difference arises from the FPV coupling, which preferentially samples flamelet states close to equilibrium, where the local volumetric heat-release rate is lower than for strained burning flamelets. The FPV reaction zone is also distributed over a broader spatial region because the returned source term depends not only on the mixture-fraction field, but also on the spatial distribution of the flamelet parameter $\lambda$. This broader reaction zone explains why the FPV model produces heat addition and maximum temperatures comparable to those of the $\epsilon$--$Z$ model despite its lower peak heat-release rate. The spatial distribution of $\lambda$ also contributes to spurious secondary and tertiary heat-release regions predicted by the FPV model near the splitter-plate trailing edge, as indicated by the light blue regions surrounding either side of the primary reaction zone in panels b) and f). These regions occur where $\lambda$ departs from its maximum value and reaches values corresponding to high-strain-rate flamelet states, even though the local mixture state is not close to the stoichiometric mixture-fraction surface. Similarly, the initial displacement of the FPV reaction zone toward the air side of the mixing layer may be attributed, at least in part, to the dependence of the FPV source term on both the mixture-fraction field and the transported progress-variable field through the inversion of $\widetilde{C}(\lambda)$.

For the $\epsilon$--$Z$ model, the white contour lines in panels c) and h) indicate the location of local flammability-limit strain rate, $S^*_{\mathrm{fl}}(\bar{p})$. Upstream of these isolines, the imposed flamelet strain rate remains below the flammability limit, and the flamelet can remain in a burning state. Downstream of the contour, the local strain rate exceeds the flammability limit and the flamelet is quenched. This behavior is evident immediately downstream of the splitter plate, where the high-shear region produces local extinction and the flame standoff observed in Fig.~\ref{fig:tempcont}. A second quenching region appears near $x \approx 110$ mm, where the local pressure reduction lowers the flammability-limit strain rate, consistent with the pressure-dependent limits shown in Fig.~\ref{fig:tmax_sstar} and with the temperature field in Fig.~\ref{fig:tempcont}. The flame standoff also permits oxidizer entrainment into the fuel side before ignition, as indicated by the red $\widetilde{Y}_{\mathrm{O_2}}$ isolines overlaid in panel f). This entrained oxidizer supports a secondary off-manifold reaction zone along the vicinity of the $S^*_{\mathrm{fl}}(\bar{p})$ contour. Unlike in the FPV model, where the secondary and tertiary reaction zones arise from the spatial distribution of the progress variable, the reaction zone here is permitted by the availability of both reactants in accordance with the off-manifold treatment described in Sec.~\ref{sec:off_manifold}. Farther downstream, however, the entrained oxygen becomes increasingly diluted on the fuel side. As a result, the availability factor associated with the tabulated oxidizer mass fraction, $\widetilde{Y}_{\mathrm{ox}}/\widetilde{Y}_{\mathrm{ox,TAB}}$, becomes too small to sustain the tabulated reaction rate, and the off-manifold scaling suppresses the chemical source terms.

It is important to emphasize that, in comparing the $\epsilon$--$Z$ flamelet model with the OSK model, the earlier ignition predicted by OSK is not being attributed simply to faster chemistry. Rather, the distinction arises from the different closures used for turbulence--chemistry interaction. The $\epsilon$--$Z$ flamelet formulation treats combustion through a subgrid chemistry model, in which the source terms are conditioned on a strain-rate measure inferred from the modeled turbulent dissipation rate, $\epsilon$. This allows the chemical source terms to be suppressed in regions where the local turbulent time scale is sufficiently short. In contrast, the OSK model evaluates the chemical source terms directly from the mean thermochemical quantities.

\begin{figure}[h!]
\centering
\begin{subfigure}{\textwidth}
  \centering
  \includegraphics[width=1.0\textwidth]{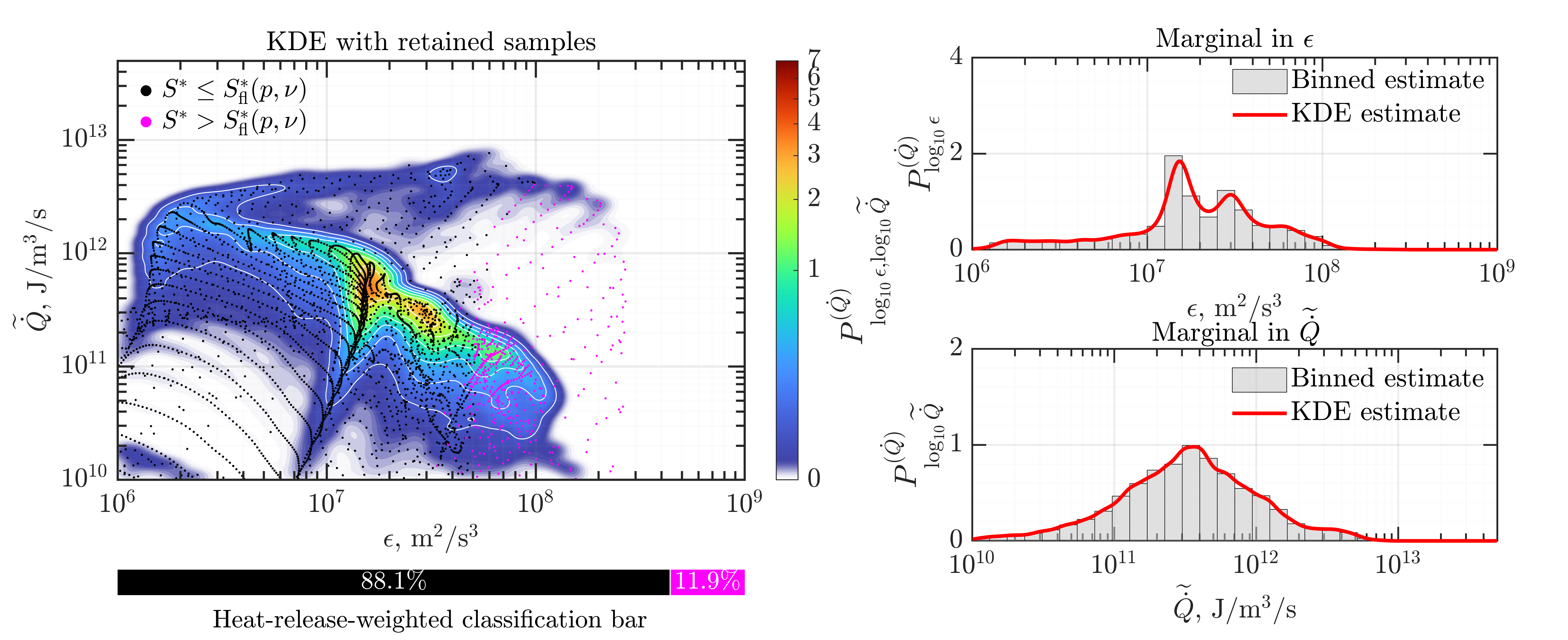}
  \caption{FPV model.}
  \label{fig:fpv_kde}
\end{subfigure}
\begin{subfigure}{\textwidth}
  \centering
  \includegraphics[width=1.0\textwidth]{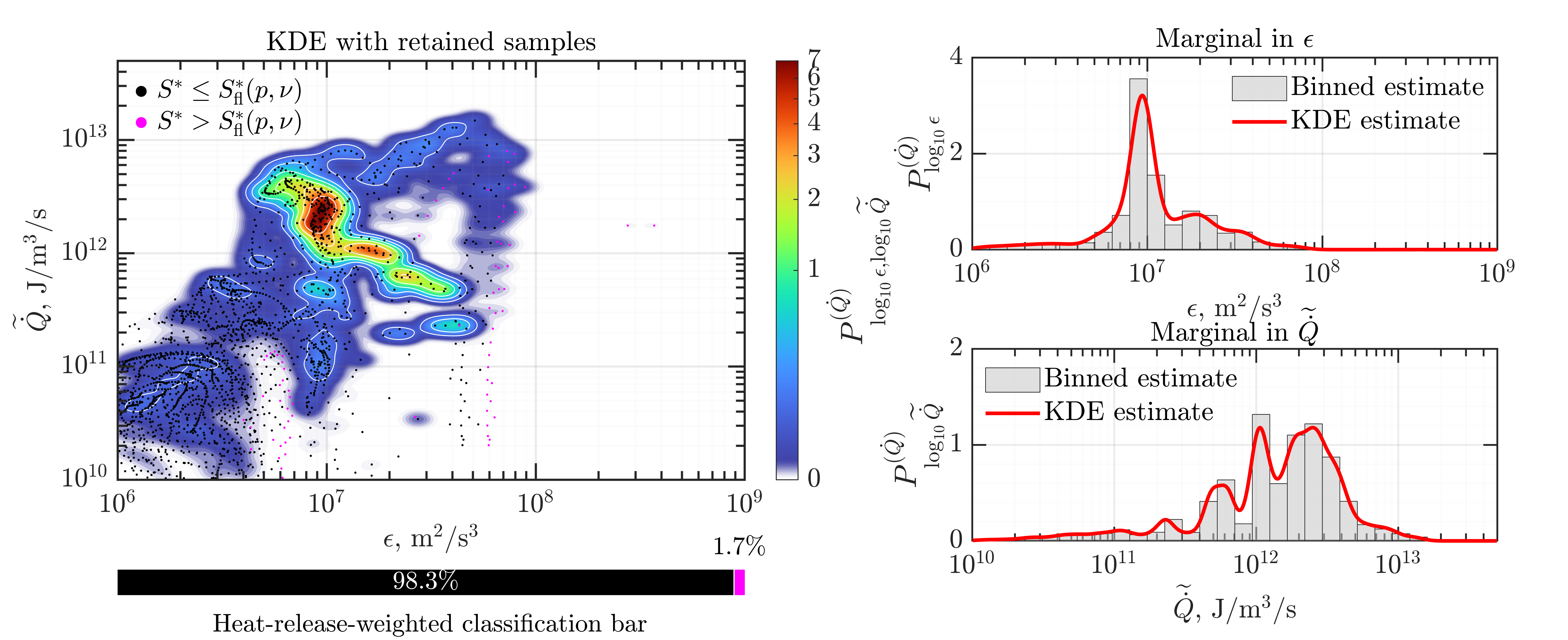}
  \caption{$\boldsymbol{\epsilon}$--$Z$ model.}
  \label{fig:eps_kde}
\end{subfigure}

\caption{Heat-release-rate weighted joint statistics of ${\epsilon}$ and $\widetilde{\dot{{Q}}}$ for both flamelet models.}
\label{fig:PDF}
\end{figure}

Figure~\ref{fig:PDF} presents heat-release-weighted joint PDFs of $\epsilon$ and $\widetilde{\dot Q}$, together with the corresponding marginal PDFs, for the FPV and $\epsilon$--$Z$ models. The FPV results are shown in Fig.~\ref{fig:fpv_kde}, while the $\epsilon$--$Z$ results are shown in Fig.~\ref{fig:eps_kde}. The statistics are constructed from the computational grid cells, where each sample $i$ has local values $\epsilon_i$ and $\widetilde{\dot Q}_i$ and cell area $A_i$. Only samples satisfying $\widetilde{\dot Q}_i \geq 10^{10}\,\mathrm{J/m^3/s}$ are retained, so that the distributions are restricted to the reacting region and are not dominated by the large number of weakly reacting or non-reacting cells. The samples are weighted by their contribution to the integrated heat-release rate,
\begin{equation}\label{eq:pdf_weights}
w_i = A_i \widetilde{\dot Q}_i ,
\end{equation}
so that the resulting PDF represents the fraction of retained heat release, rather than the fraction of grid-cell area, associated with each region of $(\epsilon,\widetilde{\dot Q})$ space.

Since both $\epsilon$ and $\widetilde{\dot Q}$ span several orders of magnitude, the density is plotted per logarithmic decade, consistent with the logarithmic axes. Defining $u=\log_{10}\epsilon$ and $v=\log_{10}\widetilde{\dot Q}$, the plotted joint PDF is
\begin{equation}
P^{(\dot Q)}_{\log_{10}\epsilon,\log_{10}\widetilde{\dot Q}}(u,v)
=
(\ln 10)^2 \epsilon \widetilde{\dot Q}\,
P
(\epsilon,\widetilde{\dot Q}) .
\end{equation}
Thus, $P^{(\dot Q)}_{\log_{10}\epsilon,\log_{10}\widetilde{\dot Q}}\,du\,dv$ is the fraction of retained integrated heat-release rate contained in the logarithmic interval $du\,dv$. The marginal PDFs shown on the right are
\begin{equation}
P^{(\dot Q)}_{\log_{10}\epsilon}(u)
=
\int_{v_{\min}}^{v_{\max}}
P^{(\dot Q)}_{\log_{10}\epsilon,\log_{10}\widetilde{\dot Q}}(u,v)\,\rm{d}v,
\qquad
P^{(\dot Q)}_{\log_{10}\widetilde{\dot Q}}(v)
=
\int_{u_{\min}}^{u_{\max}}
P^{(\dot Q)}_{\log_{10}\epsilon,\log_{10}\widetilde{\dot Q}}(u,v)\,\rm{d}u .
\label{eq:marginal_pdfs}
\end{equation}
Here $u_{\min}$ and $u_{\max}$ are the minimum and maximum values of $\log_{10}\epsilon$ among the retained samples, while $v_{\min}=10$ and $v_{\max}$ is the maximum retained value of $\log_{10}\widetilde{\dot Q}$. The color contours and red marginal curves are kernel-density estimates (KDEs) of the heat-release-weighted PDFs, while the gray bars show the corresponding binned estimates obtained from the retained computational samples. The KDE colormap is nonlinearly scaled using an exponential color mapping to enhance the visibility of low-probability regions while compressing the highest-density values. Black and magenta points denote retained samples that are locally below and above the flammability limit, respectively, as determined from $S^*$ relative to the local limit $S_{\mathrm{fl}}^*(\bar{p})$. The percentages reported below the joint PDFs denote heat-release-rate weighted fractions of the retained integrated heat-release rate in each category, rather than the percentage of computational samples.

The statistics reveal that the $\epsilon$--$Z$ model localizes the dominant heat-release contribution within a relatively narrow range of dissipation rate, centered near $\epsilon=O(10^7),\mathrm{m^2/s^3}$. Since $\epsilon$ determines the Kolmogorov time and length scales through $\tau_\eta \sim (\nu/\epsilon)^{1/2}$ and $\eta \sim (\nu^3/\epsilon)^{1/4}$, this concentration indicates that the dominant heat-release rate is associated with a restricted range of small-scale turbulent states. In contrast, the FPV model distributes heat release over a broader range of $\epsilon$, indicating a weaker localization of combustion with respect to the local small-scale turbulent state. The marginal distributions of $\widetilde{\dot Q}$ further show that the two models differ not only in where heat-release rate occurs in $\epsilon$ space, but also in the intensity of the local heat-release rates that contribute to the total heat addition. The FPV model derives a larger fraction of its heat-release-weighted contribution from lower volumetric heat-release rates, whereas the $\epsilon$--$Z$ model places more of the retained heat-release rate at larger $\widetilde{\dot Q}$. This suggests that FPV produces a more spatially distributed, lower-intensity burning mode, while the $\epsilon$--$Z$ model concentrates the heat addition into a smaller subset of more intense burning states. This is consistent with the saturation of near equilibrium flamelet states selected in the FPV model.

The flammability classification provides an additional distinction between the two closures. $11.9\%$ of the retained integrated heat-release rate in the FPV model is contributed by points for which the local strain-rate estimate exceeds the pressure-dependent flammability limit, $S^* > S^*_{\mathrm{fl}}(\bar{p})$. Since this percentage is heat-release-rate weighted (with analogous definition to Eq. (\ref{eq:pdf_weights})), it does not merely indicate that some samples lie outside the nominal flammability range; rather, it shows that a non-negligible fraction of the modeled heat addition is produced by states that would be classified as locally extinguished under the strain-based criterion. The corresponding fraction for the $\epsilon$--$Z$ model is only $1.7\%$, indicating substantially better consistency between the modeled heat-release rate and the local flammability constraint. This small residual contribution is likely associated with points near the flammability boundary and is therefore sensitive to the use of a sharp cutoff in $S^*_{\mathrm{fl}}(\bar{p})$.

\subsection{Species Transport and Off-Manifold Composition}
\begin{figure}
\centering
\begin{subfigure}{.5\textwidth}
  \centering
  \includegraphics[width=0.95\textwidth]{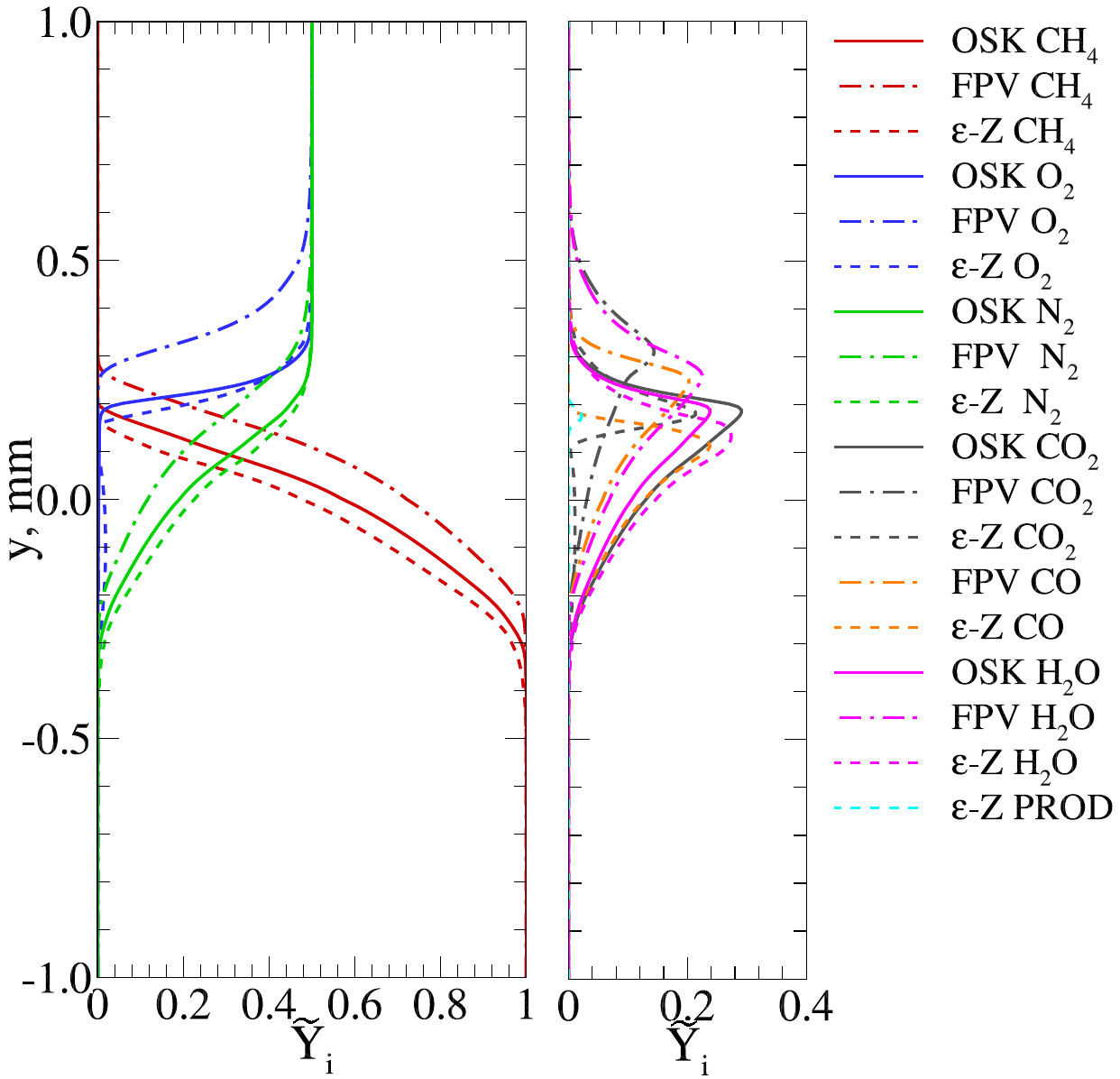}
  \caption{Species profiles at x = 25 mm.}
  \label{fig:y_profiles}
\end{subfigure}%
\begin{subfigure}{.5\textwidth}
  \centering
  \includegraphics[width=1.0\textwidth]{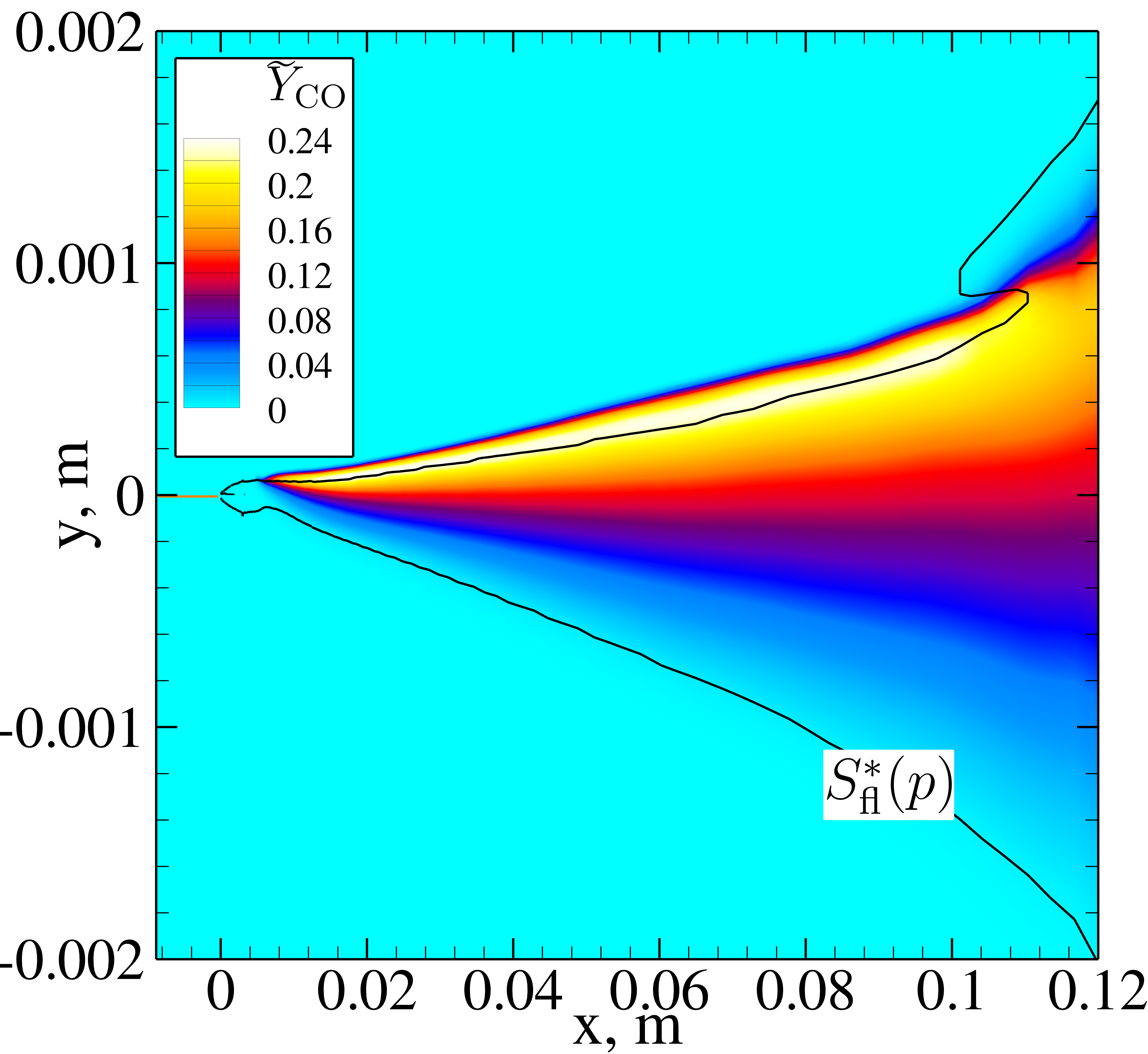}
  \caption{Contour of ${\widetilde{Y}_{\mathrm{CO}}}$ for the $\boldsymbol{\epsilon}$-based results.}
  \label{fig:y_co}
\end{subfigure}
\caption{Mixture compositions and ${\widetilde{Y}_{\mathrm{CO}}}$ field.}
\label{fig:mass_fractions}
\end{figure}

Accounting for species transport through explicit species conservation equations, as done in the OSK and $\epsilon$--$Z$ models, produces noticeably different mixture compositions from those obtained by retrieving species mass fractions directly from a precomputed flamelet table, as in the FPV model. Figure~\ref{fig:y_profiles} shows the mixture-composition profiles at $x$ = 50 mm for all three models. Both flamelet models differ substantially from the OSK model, which is expected because the flamelet formulations use a multistep chemical mechanism, whereas OSK employs a one-step global reaction model. Consequently, differences in the product-species distributions should not be attributed solely to transport effects, but also to the different levels of chemical detail in the models. The FPV model predicts broader species profiles, consistent with the wider thermal and velocity mixing layers observed previously. However, broadening alone does not fully explain the differences in profile shape relative to the $\epsilon$--$Z$ model. The profiles are not simply stretched versions of one another; rather, the relative shapes, gradients, and product-species distributions differ between the two flamelet closures. In particular, the $\epsilon$--$Z$ model produces sharper composition gradients and a more localized product region, consistent with the more localized heat-release field and heat-release-weighted statistics discussed previously. This indicates that the two flamelet models recover different local thermochemical states, not merely different mixing-layer thicknesses.

In the same way that the distribution of $\lambda$ affects the FPV heat-release field, it also affects the retrieved species mass fractions and the corresponding implied chemical state represented by the table. By contrast, the OSK and $\epsilon$--$Z$ models solve explicit species transport equations. Molecular and turbulent transport of individual species can therefore contribute directly to the resulting profile shapes, whereas in the FPV model these effects enter indirectly through the transported variables $\widetilde{Z}$, $\widetilde{Z''^2}$, and $\widetilde{C}$. The differences observed in Figure~\ref{fig:y_profiles} therefore suggests that explicit species transport can preserve species-composition gradients more consistently with the OSK baseline than transport through the reduced FPV variables alone. This improvement may justify the additional computational cost associated with solving species transport equations. 

% Another feature of the $\epsilon$--$Z$ profiles is the lumped product species, $\mathrm{PROD}$. Its mass fraction remains small and bounded by the imposed subgrid-composition constraint $\gamma \ge 0.95$ in Eq. (\ref{eq:prod}), corresponding to $Y_{\mathrm{PROD}} \le 5\%$. This indicates that the lumped product correction remains limited and does not dominate the resolved mixture composition.

Figure~\ref{fig:y_co} shows contours of $Y_{\mathrm{CO}}$ for the $\epsilon$--$Z$ model. The overlaid black contour line denotes the local flammability-limit strain rate, $S^*_{\mathrm{fl}}(\bar{p})$. Despite local flamelet quenching at $x = 110$ mm, where reaction rates drop to zero, the explicit species transport allows CO produced upstream to advect and diffuse downstream of the quenched region, thereby allowing for off-manifold representation. This behavior mitigates a key limitation of strain-rate-based formulations, in which direct interpolation of species mass fractions from flamelet tables leads to abrupt transitions between burning and non-burning states along the S-curve. The FPV model introduces the progress variable specifically to smooth these discontinuities by providing a continuous mapping between flamelet branches. However, our results show that when species transport is explicitly solved on the resolved scale, the species concentrations remain continuous across the quenched region without the need for a progress variable, while only the chemical source terms exhibit discontinuities.